\documentclass[aps, prx,twocolumn,superscriptaddress,showpacs,amsmath,amssymb]{revtex4-1}

\usepackage{amssymb}
\usepackage[dvips]{epsfig,graphics,color}
\usepackage{multirow}
\usepackage{hhline}
\usepackage{graphicx}

\usepackage{color}
\usepackage[normalem]{ulem}

\usepackage{color}

\newcommand{\COMMENTED}[1]{}
\newcommand{\REMARKS}[1]
{ 
{ \color{red}{\textbf{ {[#1]} }} }
}

\begin{document}

\title{Benchmark study of the two-dimensional Hubbard model with auxiliary-field quantum
	Monte Carlo method
	 }

\author{Mingpu Qin}
\affiliation{Department of Physics, College of William and Mary, Williamsburg, Virginia 23187, USA}

\author{Hao Shi}
\affiliation{Department of Physics, College of William and Mary, Williamsburg, Virginia 23187, USA}

\author{Shiwei Zhang}
\affiliation{Department of Physics, College of William and Mary, Williamsburg, Virginia 23187, USA}

\begin{abstract}
Ground state properties of the Hubbard model on a two-dimensional square lattice are studied by the auxiliary-field 
quantum Monte Carlo method. Accurate results for energy, double occupancy, effective hopping, magnetization, 
and momentum distribution  are calculated for interaction strengths of $U/t$ from $2$ to $8$, for a range of densities
including half-filling and $n = 0.3, 0.5, 0.6$, $0.75$, and $0.875$. At half-filling, the results are numerically exact.
Away from half-filling, the constrained path Monte Carlo method is employed to control the sign problem. 
Our results are obtained with several advances in the computational algorithm, which are described in detail. 
We discuss the advantages of generalized Hartree-Fock trial wave functions and its connection to pairing wave functions,
as well as the interplay with different forms of Hubbard-Stratonovich decompositions. 
We study the use of different twist angle sets when applying the twist averaged boundary conditions.
We propose the use of  quasi-random sequences, which improves the convergence 
to the thermodynamic limit over pseudo-random and other sequences. 
With it and a careful finite size scaling analysis, we are able to obtain accurate 
values of ground state properties in the thermodynamic limit. Detailed results for finite-sized systems
up to $16 \times 16$ are also provided for benchmark purposes.

\end{abstract}


\maketitle

\section{Introduction}
The two-dimensional (2D) Hubbard model \cite{Hubbard_origional} is one of the simplest models which are relevant to 
many correlated electron phenomena including interaction-driven metal-insulator transitions
\cite{Imada_rmp_1998}, spin and charge density waves \cite{cdw_sdw}, 
magnetism \cite{ph-sym} and superconductivity\cite{htc}. The ability to predict
the properties of 2D Hubbard model is crucial to our understanding of the related exotic quantum states and the transition between them. Though
the one dimensional
Hubbard model is exactly solvable \cite{lieb_wu}, no exact solution for the Hubbard model exists in two or higher dimensions except for a few
special parameter values.

The ground state property of the 2D Hubbard model has been investigated by a variety of methods which have both strengths and weaknesses in different regions of
the parameter space. In a recent work \cite{paper_simons}, the 2D Hubbard model was studied by 
state-of-the-art numerical methods \cite{VMC-DMC-GFMC,MPHF,CC,DMRG,DMET,DCA,DF,diagMC}.
In the present paper, we provide a detailed account of 
the  auxiliary-field quantum Monte Carlo (AFQMC) study
 in Ref.~\cite{paper_simons}, introduce two methodological advances which improve the accuracy 
 and efficiency of AFQMC calculations, and  
 present systematic results for finite-size supercells and detailed 
analysis of the scaling to the thermodynamic limit.
Because of the high accuracy of AFQMC, the results in this paper will be able to serve as benchmarks 
for future calculations 
and method development. Such benchmarks will be very valuable given the fundamental nature of the Hubbard model.

In addition to the detailed and systematic data, we 
propose here the use of a quasi-random sequence which 
reduces the fluctuations and accelerate convergence when implementing twist averaged boundary conditions (TABC). 
We test this approach and study the convergence of different boundary conditions.
The quasi-random twist is applicable to all many-body calculations of extended systems, including realistic 
electronic structure calculations in correlated materials. 

We also describe the use of generalized Hartree-Fock (GHF) trial wave functions over 
the more standard unrestricted Hartree-Fock (UHF) form, and discuss how and when
improvement in accuracy and efficiency results, both at 
half-filling and in the doped regime.  
The connection between the GHF form for magnetic correlations (repulsive model, half-filling) and the BCS form for
superconducting order (attractive, spin-balanced model) is discussed, as well as their 
relation to the form of the many-body propagators and their symmetry properties.
Such wave functions can be readily generalized 
to other quantum Monte Carlo calculations in 
many-electron systems. 

At half-filling in the repulsive Hubbard model, 
the result from AFQMC is numerically exact and the method is 
computationally very efficient.
Away from half-filling, AFQMC methods 
suffer from the minus sign problem \cite{sign, sign_problem} associated with Fermi statistics which leads to exponentially 
growing statistical errors with system size and inverse temperature. 
We employ the constrained path formalism under AFQMC, commonly referred to as constrained path
Monte Carlo (CPMC), to control the sign problem by introducing a trial wave-function to guide the walk in 
the Slater determinant space. This restores the algebraic computational scaling as in the half-filled case, but 
introduces a possible systematic error. The goal of considering different forms of the trial wave 
function is to minimize this error, and to improve the prefactor in the algebraic scaling.

The rest of the paper is organized as follows. In Sec.~\ref{model_method}, we first define the Hubbard model
 and give a brief summary of the method used in this work. We also introduce the use of twist boundary conditions in computations of finite supercells.
 In Sec.~\ref{method_new} we describe the computational algorithmic advances. 
 The use of a quasi-random sequence in the 
 twist averaged boundary conditions is discussed, with test results presented.
 We also study the use of GHF trial wave functions and analyze their connection to BCS wave functions.
 The interplay between the form of the trial wave function and symmetry properties of the Hubbard-Stratonovich 
 transformation is examined.
In Sec.~\ref{results_half_filling}  we present detailed, exact numerical finite-size results 
at half-filling  for a range of supercell sizes and boundary conditions, from weak to strong-coupling regimes. 
A careful finite-size scaling analysis is carried out to extrapolate the computed quantities to the thermodynamic limit.
 In Sec.~\ref{result_away_half-filling}, the results for system away from half-filling are presented. A short summary
 in Sec.~\ref{conclusion} will conclude this paper.
The Appendix contain the finite size numerical data including ground state energy, double occupancy and kinetic energy.

\section{Model and Method}
\label{model_method}
\subsection{Hubbard Model}
The Hubbard model is defined as
\begin{equation}
H = K + V = -\sum\limits_{ i,j, s}t_{ij}  \left(c_{i,s}^\dagger c_{j,s}  + H.c.\right) +U\sum\limits_i n_{i\uparrow}n_{i\downarrow},
\label{eqn:H}
\end{equation}
where $K$ and $V$ are the kinetic and interaction terms, respectively. 
\COMMENTED{
\REMARKS{I changed spin symbol from $\sigma$ to $s$ to remove inconsistency with def in spin corr later.
Check to make sure we didn't miss any spots}
}
The creation (annihilation) operator on site $i$ is
$c_{i,s}^\dagger$ ($c_{i,s}$), with $s = \uparrow,\downarrow$ the spin
of the electron,
and $n_{i,s}$ is the corresponding number operator.
We denote the total number of electrons with up and down spin by
 $N_\uparrow$ and $N_\downarrow$. In this work, we only consider the spin-balanced ($N_\uparrow = N_\downarrow$) systems. The filling factor is defined as $n = (N_{\uparrow}+N_{\downarrow} )/{N}$ where
$N$ is the total  number of lattice sites in the supercell. Half-filling is $n=1$, and away from it the hole density is given by 
$h = 1 -n$.
We deal with only nearest neighboring and uniform hopping, $t_{ij}=t$ for each 
near-neighbor pair $\langle ij\rangle$, and set $t$ as the energy unit. 
The strength of the repulsive interaction is given by $U/t$.
With the exception of 
the $h = 1/8$ doping case where rectangular lattice are studied to accommodate the underlying spin
density wave structure, we consider supercells of square lattice with size $N = L \times L$.

In order to better extrapolate to the  thermodynamic limit (TDL), we use  
TABC \cite{TBC}.
As shown in Sec.~\ref{results_half_filling}, the standard periodic boundary conditions (PBC) turns out to give 
non-monotonic convergence with supercell size.
Under twist boundary conditions (TBC), an electron
gains a phase when hopping across the boundaries:
\begin{equation}
  \Psi(\ldots,\mathbf{r}_j+\mathbf{L},\ldots) = e^{i\hat{\mathbf{L}}\cdot\mathbf{\Theta}}\Psi(\ldots,\mathbf{r}_j,\ldots),
  \label{eqn:tbc}
\end{equation}
where $\hat{\mathbf{L}}$ is the unit vector along $\mathbf{L}$, and
the twist angle $\mathbf{\Theta}=(\theta_x,\theta_y)$ is a
parameter, with $\theta_x$  ($\theta_y$)  $\in [0, 2\pi)$. 
This is equivalent to placing the lattice on a torus topology and applying a 
magnetic field which induces a flux of $\theta_x$ along the $x$-direction (and a flux of $\theta_y$ along the $y$-direction).
In Eq.~(\ref{eqn:tbc}), the 
translational symmetry is explicitly broken, but we can also choose another gauge with which the translational symmetry is preserved, i.e., adjust $t$ to
$t \times e^{i\theta_x / L}$ along $x$ and $t \times e^{i\theta_y / L}$ along $y$. 
By imposing a random TBC, the possible degeneracy of the non-interacting energy levels
is lifted by breaking the rotational symmetry of the lattice. This eliminates 
the so called open-shell effects.
 
 To implement TABC, we choose a set of  $N_\theta$  twist angles
 and carry out the calculation for each separately. 
 The constrained path condition can be generalized straightforwardly to the case of TBC \cite{chia-chen_EOS}.
The computational cost is thus nominally $N_\theta$ times that of a single calculation for, say, the PBC.
However, by averaging 
 the same physical quantities from all the calculations, 
 the statistical error bar of the TABC value of the given quantity is reduced.
 As will be discussed later, the associated statistical uncertainty can be estimated
 from the distribution among the twist angles. 
For non-interacting systems, the TABC energy at half-filling approaches the exact TDL value as $N_\theta$ is increased.
However, if the canonical ensemble is used with fixed particle number $N$, the TABC result with $N_\theta \rightarrow \infty$ 
is in general not equal to the TDL value \cite{cheong_thesis,TBC}. This is the case away from half-filling in the 2D Hubbard model.
 (Of course the discrepancy goes to zero as 
the system size $N$ is increased.)

 The use of TABC and the treatment of 
 finite-size effects, including the effect from electron correlations,  have been discussed earlier \cite{Chiesa-FS,Hendra-FS,chia-chen_EOS}.
 The quasi-random sequence we discuss below can be directly applied
 in this framework.
 Recently another method has been proposed to reduce the one-body 
 finite-size effect in the Hubbard model by modifying the 
 energy levels of the free electron part of the Hamiltonian in a way consistent with the corresponding
 one-particle density of states in the TDL 
 \cite{sandro_prb_2015}. 
 In this work, we have chosen to treat the original Hubbard Hamiltonian, since part of our goal is to produce
 benchmark data for finite-size supercells.

\subsection{Auxiliary-field Monte Carlo method}
In this section, we will briefly introduce the AFQMC \cite{AFQMC} method. (For a comprehensive
discussion of this method, see Ref.~\cite{lecture-notes}.)  
By repeatedly applying the projection operator to a state $|\psi_0\rangle$ whose 
overlap with the ground state $|\psi_g\rangle$ of the Hamiltonian $H$ in Eq.~(\ref{eqn:H}) is nonzero, we can obtain $|\psi_g\rangle$:
\begin{equation}
|\psi_{g}\rangle \propto \lim_{\beta \rightarrow\infty}e^{-\beta H}|\psi_{0}\rangle
\label{eqn:proj}
\end{equation}
and the expectation value of an operator $O$ can be calculated as
\begin{equation}
\langle O\rangle=\frac{\langle\psi_{0}|e^{-\beta H}Oe^{-\beta H}|\psi_{0}\rangle}{\langle\psi_{0}|e^{-2\beta H}|\psi_{0}\rangle}\,.
\label{eqn:expect_O}
\end{equation}

Through the Trotter Suzuki decomposition, we can decouple the kinetic and interaction part in the projection operator:
\begin{equation}
e^{-\beta H}=(e^{-\tau H})^{n}=(e^{-\frac{1}{2}\tau K}e^{-\tau V}e^{-\frac{1}{2}\tau K})^{n}+O(\tau^{2})
\end{equation}
where $\beta = \tau n$. The Trotter error can be eliminated by an extrapolation of $\tau$ to $0$. 
We typically choose $\tau = 0.01$ in this work, with which we have verified that the Trotter error is below 
the targeted statistical errors.

We usually choose $|\psi_0\rangle$ as a Slater determinant in AFQMC. The one-body term $e^{-\frac{1}{2}\tau K}$ can be directly applied to it and 
the result is another Slater determinant. This does not hold for the two-body term $e^{-\tau V}$. However,
we can decompose the two-body term into an integral of one-body terms through the so-called Hubbard-Stratonovich (HS) transformation. There exist different
types of HS transformations for $e^{-\tau V}$. The two commonly used types in the literature are the so called spin decomposition  
\begin{equation}
e^{-{\tau}Un_{\uparrow}n_{\downarrow}}=e^{-{\tau}U(n_{\uparrow}+n_{\downarrow})/2}\sum_{x=\pm1}\frac{1}{2}e^{\gamma_s x(n_{\uparrow}-n_{\downarrow})}\,,
\label{eq:spin-decom}
\end{equation} 
with the constant $\gamma_s$ is determined by $\cosh(\gamma_s) \equiv  \exp({\tau}U/2)$, and the charge decomposition
\begin{equation}
e^{-{\tau}Un_{\uparrow}n_{\downarrow}}=e^{-{\tau}U(n_{\uparrow}+n_{\downarrow}-1)/2}\sum_{x=\pm1}\frac{1}{2}e^{\gamma_c x(n_{\uparrow}+n_{\downarrow}-1)}\,,
\label{eq:charge-decomp}
\end{equation}
with
$\cosh(\gamma_c) \equiv \exp(-{\tau}U/2)$ \cite{hirsch_prb_1983}. Here $x$ is an Ising-spin-like auxiliary field. 
Different choices of the HS can lead to different accuracies or efficiencies, because of symmetry 
considerations \cite{CPMC_sym_1,CPMC_sym_2} or other factors \cite{Hao-inf-var}. We will further comment on 
the decompositions later.

After the HS transformation,
Eq.~(\ref{eqn:expect_O}) turns into
\begin{equation}
\langle O\rangle=\frac{\sum_{\{X_{i},X_{j}\}}\langle\psi_{0}|\prod_{i=1}^{n}P_{i}(X_{i})O\prod_{j=1}^{n}P_{j}(X_{j})|\psi_{0}\rangle}{\sum_{\{X_{i},X_{j}\}}\langle\psi_{0}|\prod_{i=1}^{n}P_{i}(X_{i})\prod_{j=1}^{n}P_{j}(X_{j})|\psi_{0}\rangle}
\label{eq:obs}
\end{equation}
where $X_{i}$ is the collection of the $N$ auxiliary fields introduced by the HS transformation, and $P_i$ is the product of the kinetic term $e^{-\frac{1}{2}\tau K}$ and the one-body terms from the HS transformation at time slice $i$.
The multi-dimensional integrals 
can then be computed by 
Monte Carlo methods, e.g., with the Metropolis algorithm.

At half-filling,
the denominator of Eq.~(\ref{eq:obs}) is
always non-negative because of  particle-hole symmetry\cite{ph-sym}. Away from half-filling, the denominator of Eq.~(\ref{eq:obs}) will in general become negative for some auxiliary fields.
In this situation the direct evaluation of Eq.~(\ref{eq:obs}) by Monte Carlo will suffer from the  sign problem \cite{sign,sign_problem}.
The sign problem can be eliminated by the constrained path approximation. The framework within which 
this has been implemented in Hubbard-like model has been referred to as the constrained path 
Monte Carlo (CPMC) method \cite{zhang_prb_1997}. 
To ensure the 
denominator in Eq.~(\ref{mix_estimate}) is positive, we constrain the paths of auxiliary-fields so that the
overlap with $|\psi_T\rangle$, computed at each time slice, remain non-negative. A description of the 
CPMC method for Hubbard-like models can be found in Ref.~\cite{Huy_CPC_2014}.

In CPMC, the wave function is represented as a linear combination of a set of slater determinants 
which are 
called walkers. The evolution of wave function in the imaginary time is represented as random walks in the Slater determinant space by 
sampling the auxiliary field. Physical quantities can be calculated using the mixed estimator as
\begin{equation}
\langle O\rangle_{\rm mixed}=\frac{\sum_{k}w_{k}\langle\psi_{T}|O|\psi_{k}\rangle}{\sum_{k}w_{k}\langle\psi_{T}|\psi_{k}\rangle}\,,
\label{mix_estimate}
\end{equation}
where $|\psi_{k}\rangle$ is the $k$th walker, $w_k$ is the corresponding weight and $|\psi_{T}\rangle$ is the trial wave-function we introduced. 
The mixed estimator is used to compute the energy (and other observables which 
commute with the Hamiltonian). For observables which do not commute with Hamiltionian, the mixed estimate
is biased, and 
back propagation is applied to correct for this \cite{zhang_prb_1997, Wirawan-PRE}.

In order to remove the sign problem, the constrained path approximation
in CPMC introduces a systematic error which depends on the
trail wave-function $|\psi_T\rangle$. With the TBC, a simple generalization of constraint can be made \cite{chia-chen_EOS}.  
Previous studies have shown the systematic error is small even with a free-electron or Hartree-Fork trial wave-function \cite{chia-chen_EOS}.
We will further discuss the accuracy of CPMC and the role of the trial wave function below in Secs.~\ref{ssec:GHF} and \ref{result_away_half-filling}.

\section{Methodological developments}
\label{method_new}

\subsection{Quasi random twist angles}
\label{ssec:QRtwist}

To implement TABC, a set of twist angles need to be chosen.
If we only consider how
to minimize the one-body finite-size effect \cite{Chiesa-FS,Hendra-FS},
the problem is related to the calculation  of a two-dimensional quadrature.
In this section, we compare three choices of random twist
angles, i.e. the pseudo random (PR) sequence, quasi random (QR) sequence and uniform grid.

A quasi random sequence is also known as low-discrepancy sequence, which is a sequence with the property that for all values of $N$, its 
subsequence $x_1, \cdots, x_N$ has a low discrepancy. Low discrepancy means the proportion of points in the sequence falling into an 
arbitrary set $B$ is close to proportional to the measure of $B$. Different from a pseudo random sequence, it fills the sampling space more 
uniformly at the price of losing some randomness. In this sense, a quasi random sequence is correlated. We choose the Halton sequences\cite{halton} 
to generate our twist angles in this work.
In the uniform grid method, the $N_{\theta x} \times N_{\theta y}$ twist angles are set as
\begin{equation}
\theta_{ij}=(\frac{2\pi}{N_{\theta x}}i,\frac{2\pi}{N_{\theta y}}j)
\end{equation}
where the integers $i = 0, \cdots, N_x - 1$ and $j = 0, \cdots, N_y -1$.
For  pseudo random twists, we generate the twist $\mathbf{\Theta}$ by pseudo random number sequence. 

\begin{figure}[t]
\includegraphics[width=8.1cm]{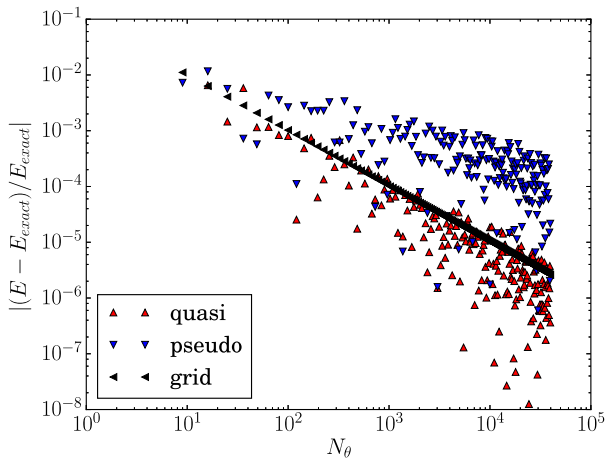}
\caption{(Color online) The convergence rate of the ground state energy of the non-interacting Hubbard model computed using TABC with 
	twist angles from QR sequence, PR sequence and uniform grid. The system is $4 \times 4$ at half-filling.
	The vertical axis shows the absolute value of the relative error with respect to the exact value, which is $-16/ \pi ^2$.}
\label{quasi_non-inter}
\end{figure}

The PR and QR twists both have residual errors which are statistical, while the grid will have a systematic
residual error. The errors vanish in the limit of a large number of twists, $N_{\theta}$. 
 From two-dimensional quadrature considerations, one would expect the
convergence rate, i.e., the residual error as a function of $N_{\theta}$, should be
$\frac{1}{\sqrt{N_{\theta}}},\frac{\ln N_{\theta}}{N_{\theta}},\frac{1}{N_{\theta}}$ 
for PR, QR, and the uniform grid, respectively.
In Fig.~\ref{quasi_non-inter} we show the convergence rates of the ground state energy of the non-interacting Hubbard model ($U = 0$) 
for the $4 \times 4$ lattice at half-filling.
The results are consistent with the expectation above. The
convergence rate with QR
TABC is almost the same as that of the uniform grid, both
much faster than with the PR sequence. 

\begin{figure}[t]
	\includegraphics[width=8.1cm]{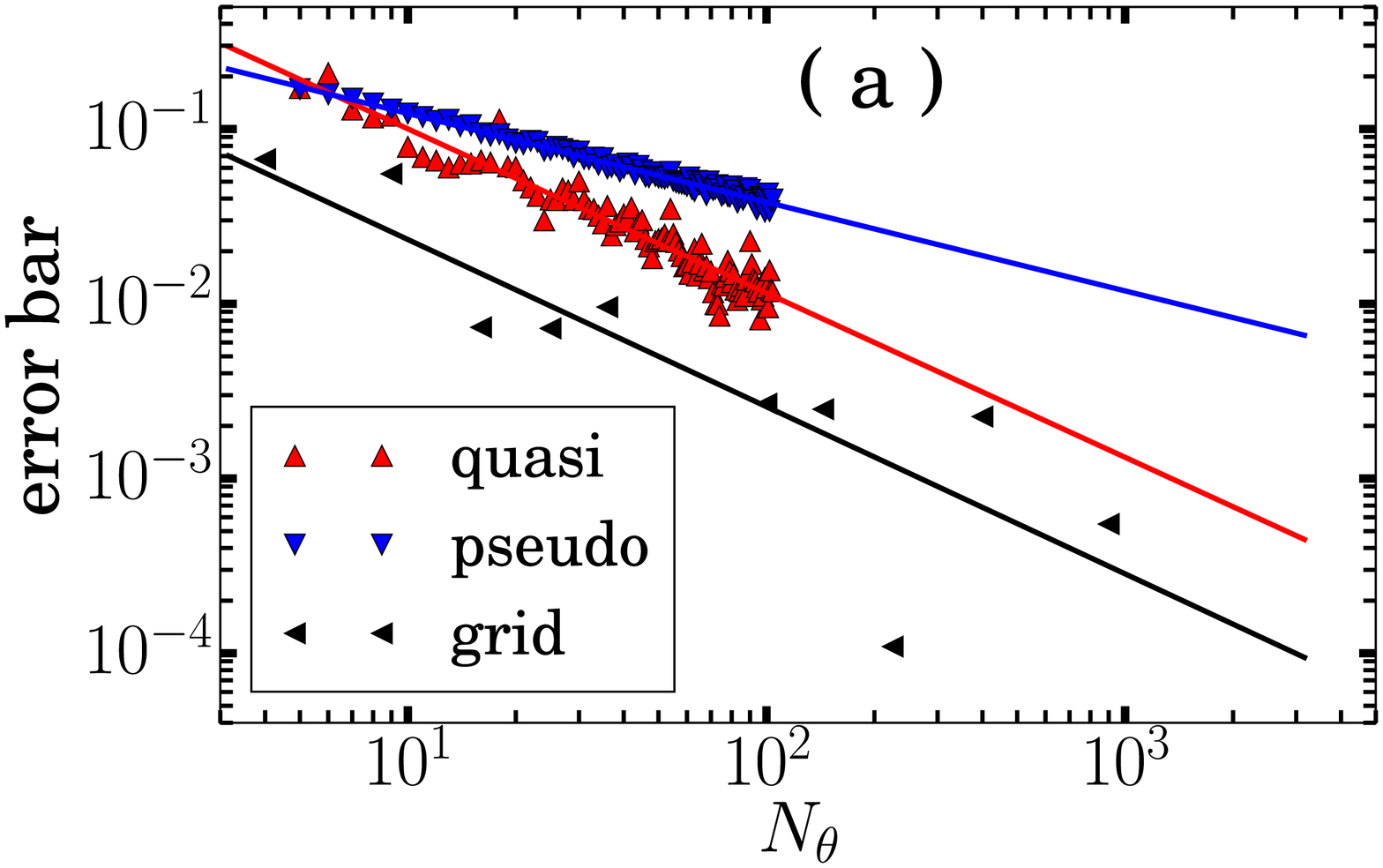}
	\includegraphics[width=8.1cm]{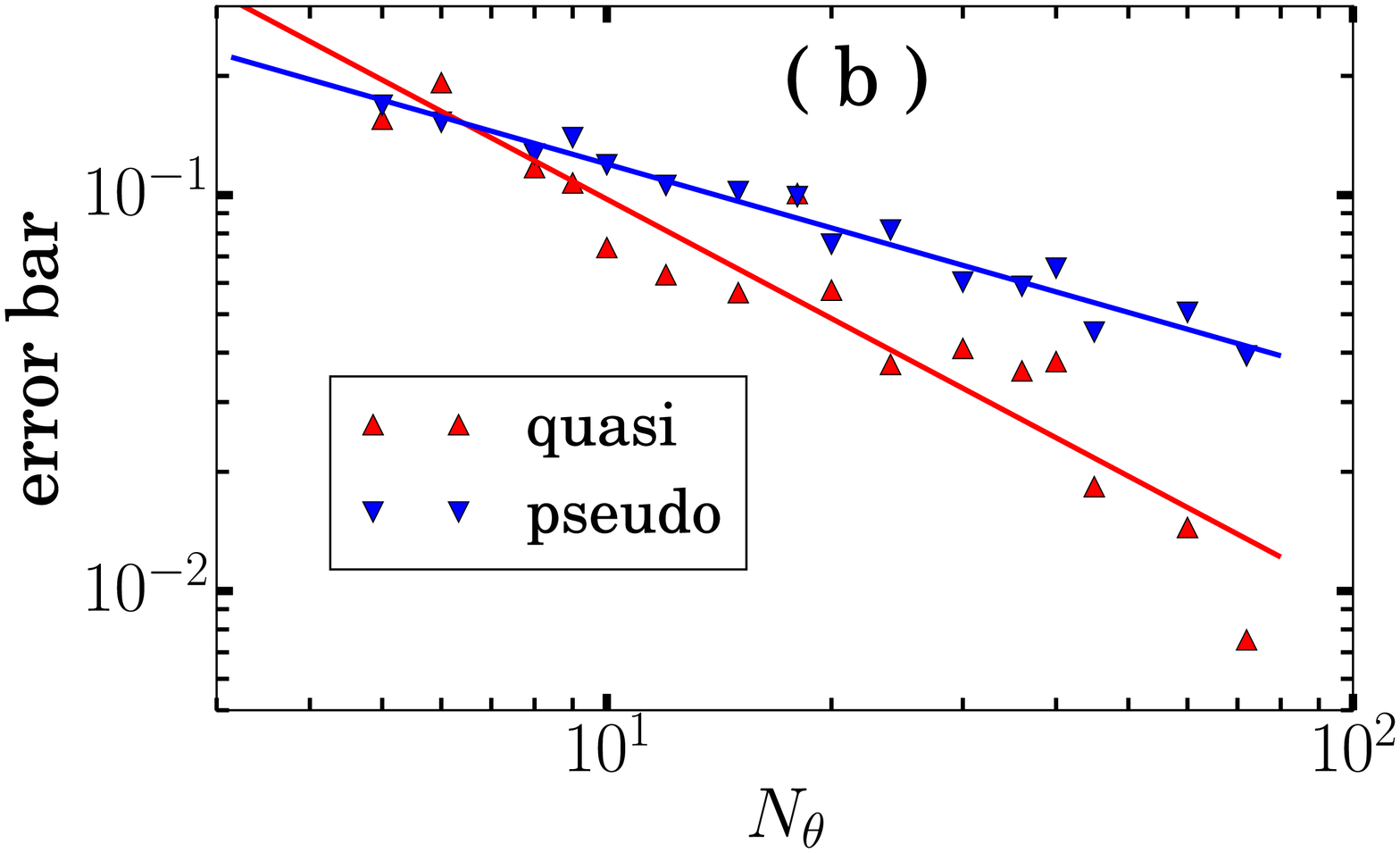}
	\caption{(Color online) (a) The error bar of the ground state energy computed 
	from TABC vs.~the number of twist angles used with twists generated by QR sequence, PR sequence, and the uniform grid. 
		The system is $4 \times 4$ with $2$ up and $2$ down electrons, and $U = 8$. The ground state energies are obtained by the ED method. 
		Note $\log$-$\log$ scale of the plot. 
		(b) Similar results for an $8 \times 8$ lattice at $n = 0.5$, with $U = 4$, for QR and PR twist angles. 
	The energies are computed by the CPMC method.
		}
	\label{quasi_inter}
\end{figure}

In Fig.~\ref{quasi_inter} (a), we study an interacting case, with $U = 8$ and a filling factor of $ n = 0.25$ 
($N_\uparrow = N_\downarrow = 2$), again in a $4\times 4$ lattice. We use exact diagonalization (ED)
to calculate the ground state energy for each twist angle.
A total of $\bar N_\theta = 3600$ twist angles are 
used in each method. To estimate the statistical error bar of the TABC energy for $N_\theta$ $(< 3600)$ twist angles for QR and PR
sequences, we partition all the data into blocks with size $N_\theta$. 
The standard derivation of the average energies from
the $[\bar N_\theta/N_\theta]$ blocks then provides an estimate of the desired statistical error. For the uniform grid, we calculate the relative error for each grid size as the difference between its average and that of the entire $3600$ twist angles. 
As in the non-interacting case shown in 
Fig.~\ref{quasi_non-inter}, the TABC energy using QR sequence
converges at a similar rate to 
that using uniform grid, with 
both
showing faster convergence rate than the PR sequence.
Linear fits of the logarithm of the ``error bar'' vs.~the logarithm of $N_\theta$ are performed , and are
shown in the figure. 
The slopes of the fit are  $0.94(3)$, $0.508(7)$ and $0.96(2)$, respectively for QR sequence, PR sequence, and the uniform grid.
These are consistent with the expected rate mentioned above.

In Fig.~\ref{quasi_inter} (b), we plot the result of an $8 \times 8$ system at $n = 0.5$. 
We use the CPMC method to 
compute the ground state energy for each twist in this system, which is well beyond the reach of ED. 
In the CPMC calculation, the corresponding non-interacting (i.e., free electron) wave function is used as a trial
 wave-function.
For many high symmetry points on a uniform grid of twist angles, the ground state of non-interacting system is degenerate. In such situations, the trial wave function (of a single Slater determinant) is not unique, 
and an arbitrary choice without consideration of symmetry properties can affect 
the accuracy of CPMC result. (This issue is further discussed below.)
To keep the analysis simple here, we only test the PR and QR sequences.
 A total of $360$ twist angles are used for both methods. 
 The same error analysis procedure is employed as in Fig.~\ref{quasi_inter} (a).
 The fitted convergence rate for PR and QR twist angles are
  $0.54(3)$ and $1.0(1)$, respectively,
again consistent with the theoretical values.

These examples show that, with QR twist angles,  the computed total energy 
from TABC converges essentially as quickly as with a uniform grid, and is much faster than with PR twists.
The use of QR twists allows the advantage of the uniform grid, while overcoming two of the 
drawbacks of the latter in QMC calculations. 
The first drawback of a uniform grid is the degeneracy which often exists with a
high symmetry grid point. As mentioned above, the degeneracy can affect the
non-interacting wave function, and correspondingly the quality of the CPMC calculation.
(Multi-determinant trial wave functions can improve the quality but they require extra handling
computationally.)
The second disadvantage of the 
uniform grid is that
one needs to determine the size of the grid prior to the calculations.
We often cannot re-use the results from a small grid size if a larger grid turns out to be necessary for 
convergence. 
On the other hand, QR sequences are cumulative.
Given that in QMC one has both statistical and convergence errors present, it is desirable 
to be able to add additional twist angles ``on the fly'' as we accumulate better indications of the magnitude of the associated errors. The QR TABC makes this possible: one can add QR twist one by one until a desired
accuracy is reached.

\subsection{GHF trial wave functions in AFQMC, and their connections to BCS wave functions}
\label{ssec:GHF}

When the sign problem is present, we use a trial wave function (TWF) to constrain the random walk paths 
in AFQMC. The sign or phase of the overlap of the sampled Slater determinants with the TWF is 
evaluated in each step, and this is used as a gauge condition which determines or modifies 
the acceptance of the 
move \cite{lecture-notes,zhang_prb_1997}. The constraint eliminates the sign or phase instability and restores the 
low-power (third power of system size here) computational scaling, at the cost of introducing, in most cases, a systematic bias. 
The quality of the TWF can  affect the accuracy of the results.
In this work we employ only single Slater determinant TWFs, which have been shown to 
 provide
accurate results in many systems. 
In Hubbard-like models, the most common choices have been the  free-electron wave function 
or the  unrestricted Hartree-Fock (UHF) solution.
The two choices each have advantages and disadvantages. The UHF is the best single Slater determinant variationally,
however, it breaks spin and translational symmetry of the system. Both symmetries, on the other hand
are preserved in the free-electron TWF. 

In this work, we use a special form of the
generalized Hartree-Fock (GHF) \cite{GHF} wave function as TWF
in the AFQMC calculations. This will increase the computation time by a factor of $2$ to $4$ in different portions
of algorithm, because now the 
Slater determinant of up and down spin are coupled. But the usage of GHF trial wave-function will reduce
the bias as shown in the discussion below.
This is 
implemented as an UHF with spin order in the $x$-$y$ plane.
As we illustrate next, this form combines the advantages of the UHF and free-electron TWFs and performs better than
both, even though it is related to the $z$-direction UHF by a spin rotation and is variationally the same.

\begin{table}[h]
\caption{The effect of the trial wave function on the (artificial) constraint and its interplay with the form of the 
HS transformation. Ground state energies are shown for $4 \times 4$ with PBC ($U = 4$) at half-filling. The exact ground
state energy is $-13.62192$. The UHF and GHF trial wave-functions are also with $U = 4$.
}
	\begin{ruledtabular}
	\begin{tabular}{cccc}
		Trial WF   & HS-spin & HS-charge \\  \hline
		UHF  &$-13.478(2)$ &  $-13.6222(2)$ \\ \hline
		GHF & $-13.623(1)$& $-13.6223(2)$ \\
	\end{tabular}
\end{ruledtabular}
\label{decomp_trial}
\end{table}

In Table~\ref{decomp_trial} we compare the effects of different TWFs and 
different HS decompositions. The system is a $4 \times 4$ with PBC and $U = 4$.
The spin and charge decompositions are defined in Eqs.~(\ref{eq:spin-decom}) and
(\ref{eq:charge-decomp}), respectively. The AFQMC results have been extrapolated to the 
$\tau = 0$ limit.
The system is at half-filling, where there is no sign problem.
The CP calculations can be easily made exact by redefining the importance sampling to have a nonzero minimum \cite{CPMC_sym_1}. However, we deliberately apply the constraint as usual, which 
can prevent 
the walkers from tunneling from one region of the determinant space to another with an artificial 
boundary where $\langle\psi_T|\psi_k\rangle=0$, even though both sides are positive. 
As can be seen, with the spin decomposition, the calculations using the UHF as a TWF leads to a bias.

When the GHF trial wave-function
is used instead of the UHF, the bias of the spin-decomposition calculation is removed. 
Below we further discuss the symmetry properties of the GHF to explain why it is a better TWF.
With the charge decomposition, the energies agree well with the 
exact energy regardless of which trial wave-function is used.  
This is because the auxiliary fields 
are complex in this case. The sign problem would become a phase problem \cite{Phaseless}. 
However, since we are at half-filling,
the overlap $\langle \psi_T|\psi_k\rangle$ turns out to be real and non-negative for all 
configurations of auxiliary-fields. 
This ``two-dimensional'' nature of the random walks \cite{Phaseless,Hao-inf-var} allows 
 ergodicity, and there is no constraint error.
 
The statistical error bars are also much smaller with the charge 
 than with the spin decomposition for the same amount of computing, as seen in Table~\ref{decomp_trial}.
 This is because the former preserves SU(2) symmetry of spin degree of freedom \cite{CPMC_sym_1}:
 when we choose as initial state for the projection
 the non-interacting  wave-function \cite{small_twist}, all the random walkers will stay in the 
 singlet space throughout the random walks, reducing fluctuations.
 (When $|\psi_T\rangle$ is used as the initial state as opposed to the non-interacting wave function, the 
equilibration time becomes much longer \cite{Wirawan-F2-spin-contamination}, and the fluctuations are larger.
The final converged results are consistent with each other  between the two different initial states, as
we would expect.)

\begin{table}[h]
\caption{Effect of the GHF trial wave function on the constraint. Mean absolute relative errors of the CP ground-state energy 
	are shown from TABC with free, UHF, and GHF trial wave functions. All UHF and GHF trial wave-functions are generated with
	effective $U$ of $4$. The system is a $4 \times 4$ lattice with
		$N_\uparrow = 7$ and $N_\downarrow = 7$. A total of $60$ twist angles are used.
		The TABC results from ED are $-16.3964$ and $-12.1510$ for $U = 4$ and $U = 8$ respectively.
	}
	\begin{ruledtabular}
	\begin{tabular}{cccc}
		        TWF &  $U = 4$ & $U = 8$ \\ \hline
		        free &  $0.51 (2) \%$ & $ 1.8 (1)\%$   \\ \hline
		        UHF  &    $0.16 (2)\%$ & $1.1 (1)\%$ \\ \hline
		        GHF  &    $0.21 (1)\%$ &  $0.51 (4)\%$ \\
	\end{tabular}
	\end{ruledtabular}
\label{4-4_7_7_GHF_UHF}
\end{table}

In Table~\ref{4-4_7_7_GHF_UHF}, we illustrate the effects 
away from half-filling. We compare the TABC energy of $4 \times 4$, $U = 4, 8$ systems at $n = 0.875$ using 
non-interacting (free), UHF, and GHF trial wave-functions. Spin decomposition are used in this case.
For simplicity, we use a uniform parameter in $z$ ($x$) direction for the UHF (GHF) calculation. In principle, we
can implement a full UHF (GHF) calculation which will improve the quality of trial wave-functions.
For $U = 4$, the result from GHF is similar to that from UHF.
Improvement with the GHF can be seen for the $U = 8$ case, with a CPMC energy closer to the exact value.

Next we further discuss the nature of the GHF wave function, its connection to 
Bardeen-Cooper-Schrieffer (BCS) wave functions \cite{BCS_wave-f}, and correspondingly, the connection between 
the repulsive Hubbard model we have studied, and the attractive model.
Let us consider a partial particle-hole transformation $\hat{P}$, which only involves
spin-$\uparrow$ electrons:
\begin{equation}
\hat{P}^\dagger c^\dagger_{i \uparrow}  \hat{P} = (-1)^i c_{i \uparrow} 
\end{equation}
and
\begin{equation}
\hat{P}^\dagger c_{i \uparrow} \hat{P} = (-1)^i  c^\dagger_{i \uparrow}.
\end{equation}
This operator $\hat{P}$ transforms the interaction term in Hubbard model from repulsive to attractive (from $U$ to $-U$) but
leaves the hopping term unchanged.

For the attractive Hubbard model,
the best mean-field description is given by the BCS theory,
\begin{equation}
\hat{H}_{BCS} = -t\sum\limits_{ \langle ij\rangle, s} \left(c_{i,s}^\dagger c_{j,s}  + H.c.\right) +\sum\limits_i \left( \Delta_i c^\dagger_{i \uparrow}  c^\dagger_{i \downarrow} + H.c.\right),
\label{eqn:bcs}
\end{equation}
where $\Delta_i$ is the order parameter.
This Hamiltonian can be transformed back to the repulsive case
 \cite{Scalettar_1989,ph-sym}.
\begin{equation}
\hat{H}_{GHF} = -t\sum\limits_{ \langle ij\rangle, s} \left(c_{i,s}^\dagger c_{j,s}  + H.c.\right) +\sum\limits_i \left( M_i c^\dagger_{i \uparrow}  c_{i \downarrow} + H.c.\right)
\label{eqn:ghf}
\end{equation}
with $M_i = (-1)^i\Delta_i$.
The
ground state of $\hat{H}_{GHF}$ is a GHF wave function, 
with antiferromagnetic order along the $x$-$y$ plane.
In other words, the GHF wave function for the repulsive model corresponds to the BCS for the 
 attractive Hubbard model. (The UHF wave function corresponds to a charge-density wave 
 restricted Hartree-Fock single Slater determinant for the attractive model.)

Symmetry properties of an AFQMC calculation directly affects its accuracy and efficiency \cite{CPMC_sym_1,CPMC_sym_2}.
The
BCS wave function conserves translational
symmetry as shown in Eq.~(\ref{eqn:bcs}), while breaking the conservation of particle numbers. 
In AFQMC calculations of an attractive Hubbard model (with $N_\uparrow=N_\downarrow$ at any 
density), the walkers will break translational symmetry because of the fluctuating auxiliary-fields, 
which are site-dependent if the charge decomposition is used. However, the walkers remain 
single determinant with fixed particle numbers. Thus the AFQMC calculation using a BCS trial 
wave function \cite{FG2D-Hao,BCS_wave-f} will have all symmetries conserved. 

With particle-hole transformation, similar arguments apply to the GHF wave function in the repulsive case. Particle-number symmetry translates to spin symmetry along the $x$-$y$ plane. 
The GHF preserves all the other symmetries except
magnetic order in the plane. When combined with UHF-type walkers which always preserve magnetic order in the plane,
all symmetries are conserved during the AFQMC calculation.

We can also think of
BCS or GHF wave functions as linear combinations of single Slater determinants.
A BCS wave function can be written as the UHF wave function plus all possible double excitations 
($ c^\dagger_{i \uparrow}  c^\dagger_{j \downarrow}$), which is a large multi-determinant wave function. 
Similarly, the GHF wave function is the UHF wave function with all possible spin-orbit excitations ($ c^\dagger_{i \uparrow}  c_{j \downarrow}$), again a multi-determinant wave function. It is thus reasonable to expect the 
GHF wave function
to perform better than the UHF.

Incidentally, since the charge decomposition is transformed to the spin decomposition
 under the particle-hole transformation, 
results in Table~\ref{decomp_trial} would indicate that spin decomposition would always give correct
results for the attractive Hubbard model.
Further, the BCS trial wave functions would have no constraint bias.
The latter is consistent with observations from calculations in Fermi gas systems in the three dimensions \cite{BCS_wave-f} and two dimensions \cite{FG2D-Hao}.

\begin{figure*}[t]
	\includegraphics[width=8.0cm]{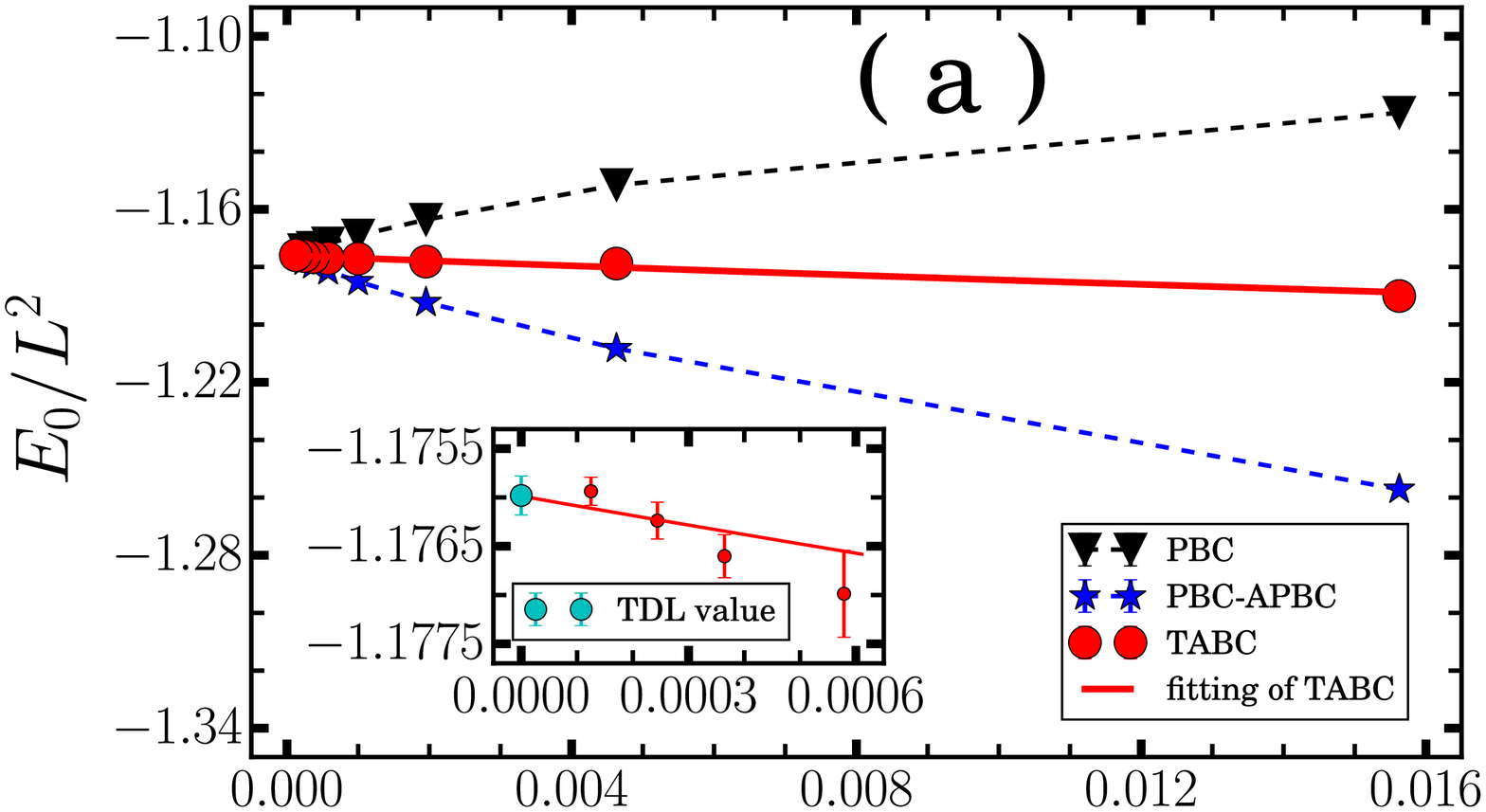}
	\includegraphics[width=8.0cm]{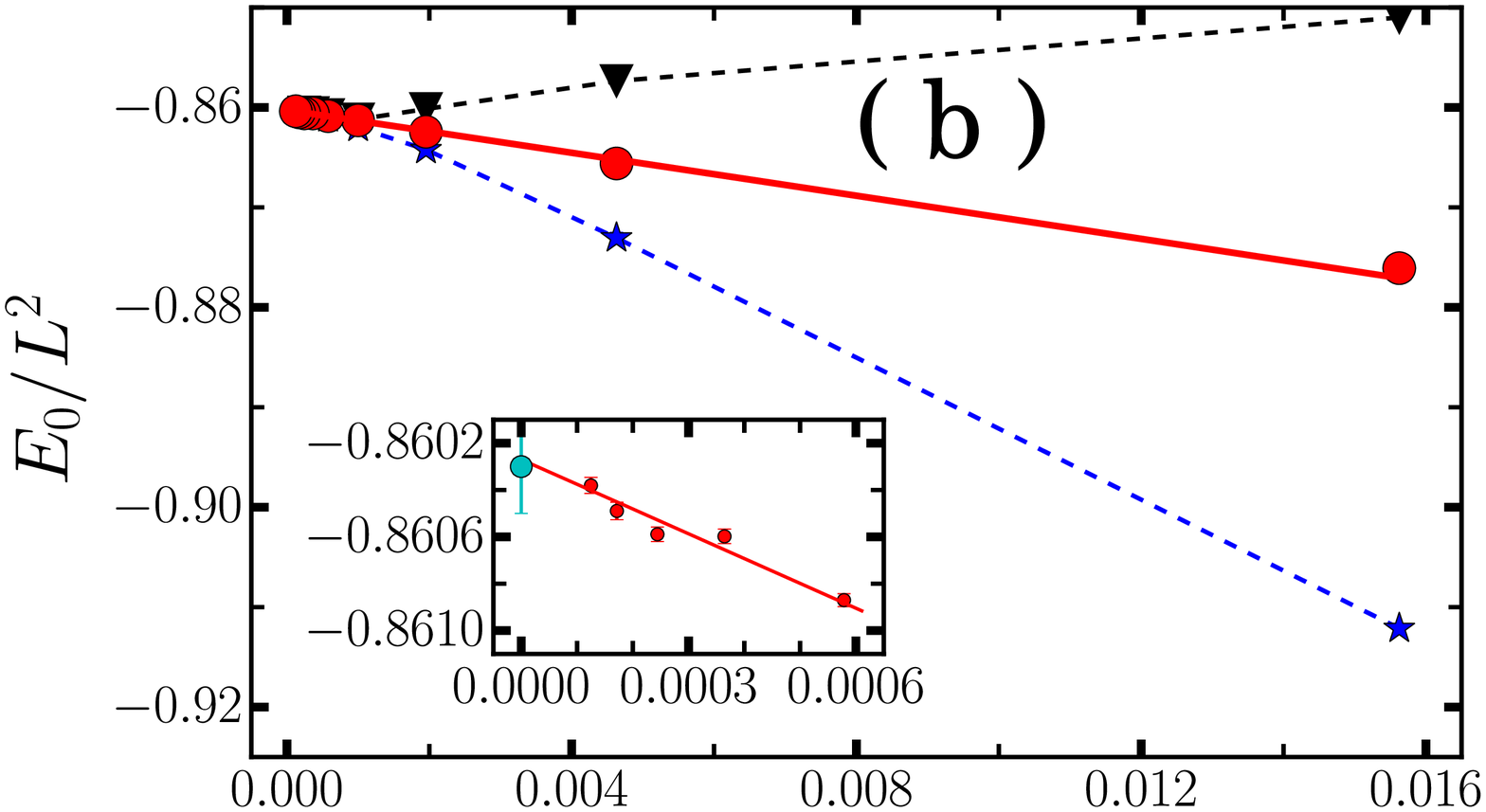}
	\includegraphics[width=8.0cm]{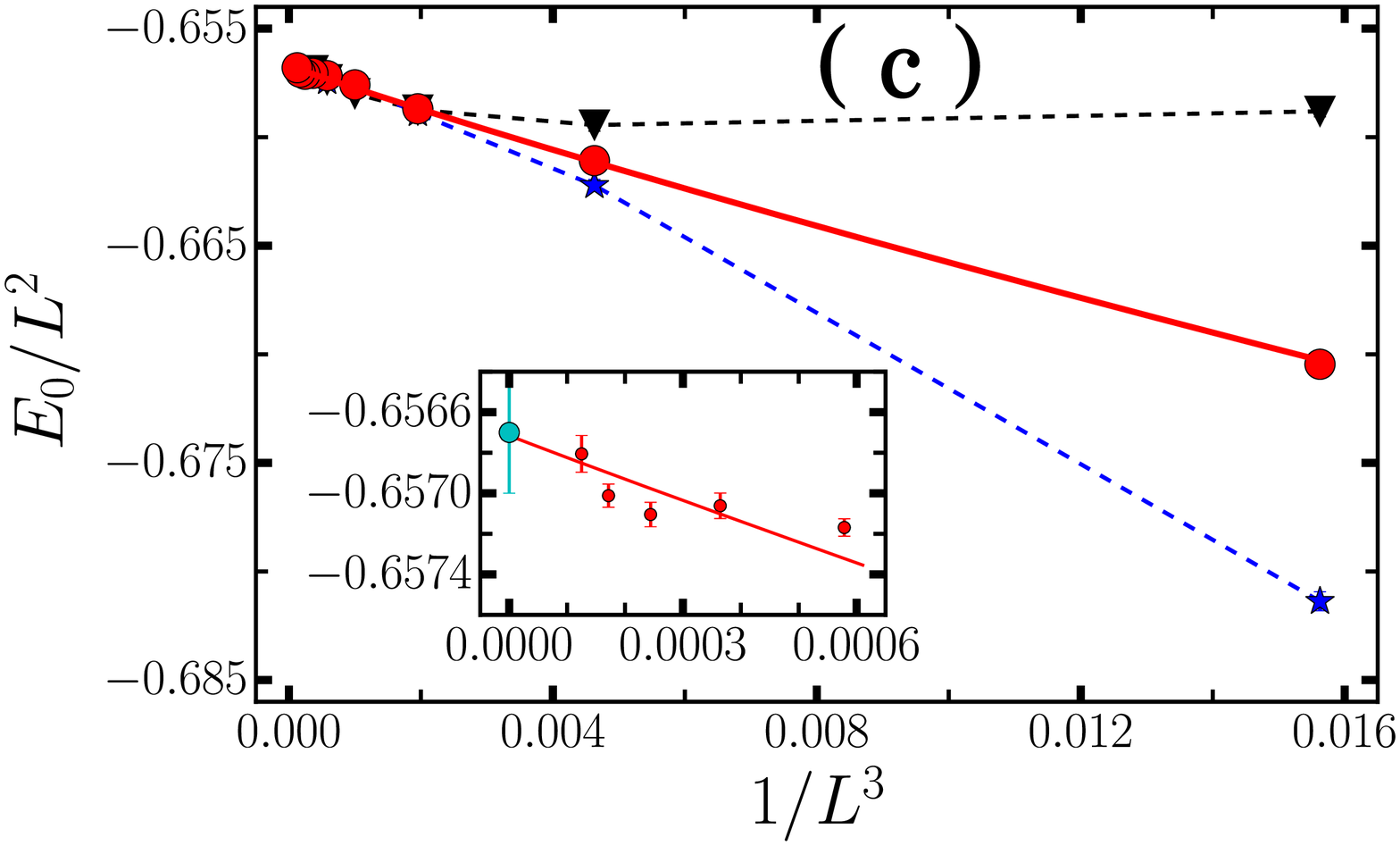}
	\includegraphics[width=8.0cm]{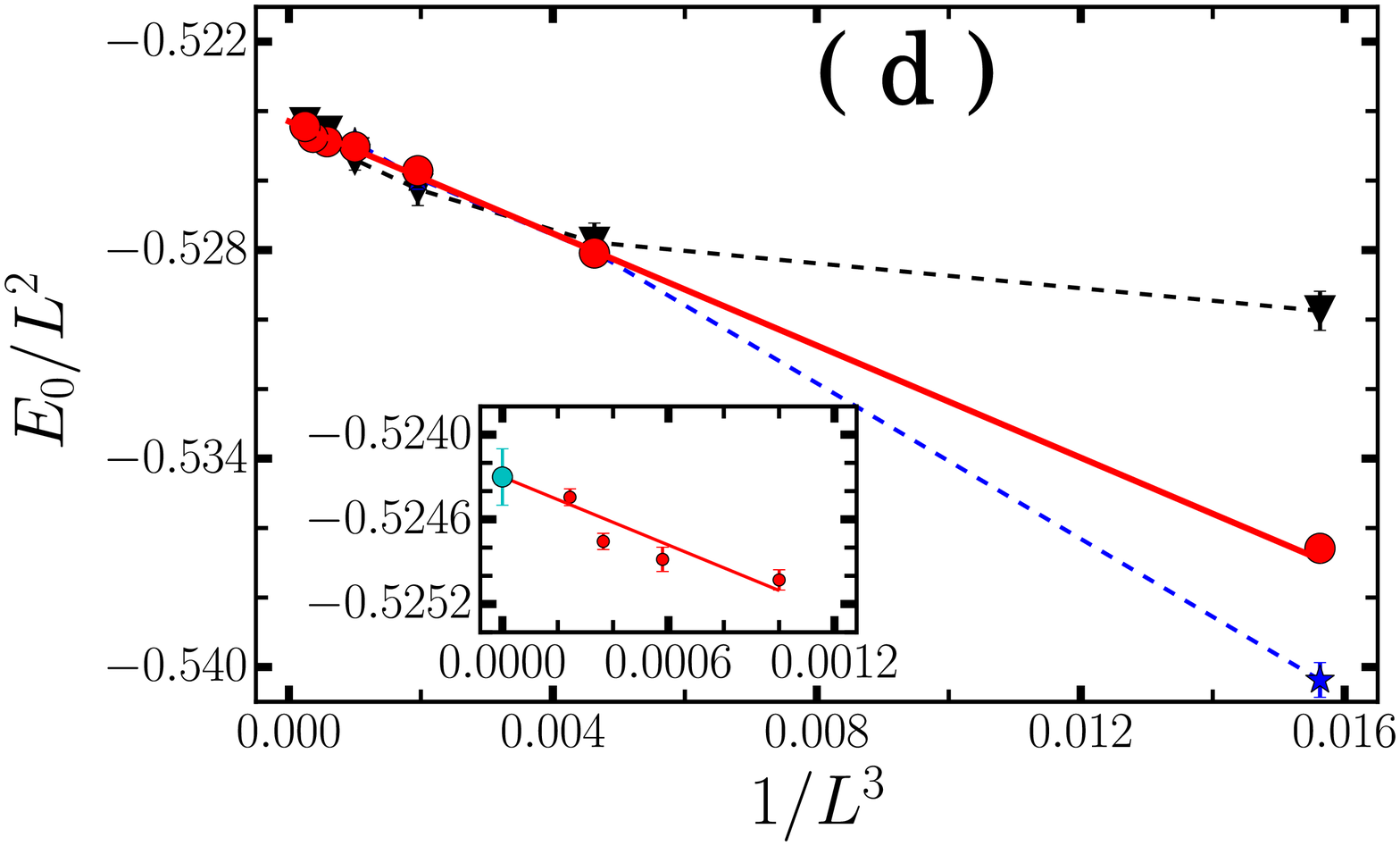}
	\caption{(Color online) Ground state energy at half-filling calculated using different boundary conditions. PBC, PBC-APBC and TABC data are represented by black triangular, blue star and red dot, respectively. 
	A fit of the TABC data is also shown, with solid red line. 
	Panels (a), (b), (c), (d) correspond to results for $U = 2, 4 , 6, 8$. In the insets of each panel, a zoom of 
	the TABC results and the fit are shown for large supercell sizes.  
	The cyan dot in each inset represents the TDL value and combined statistical and twist error bars and 
	the uncertainty from the fit. }
	\label{E_half_all}
\end{figure*}

\begin{figure*}[t]
	\includegraphics[width=8.0cm]{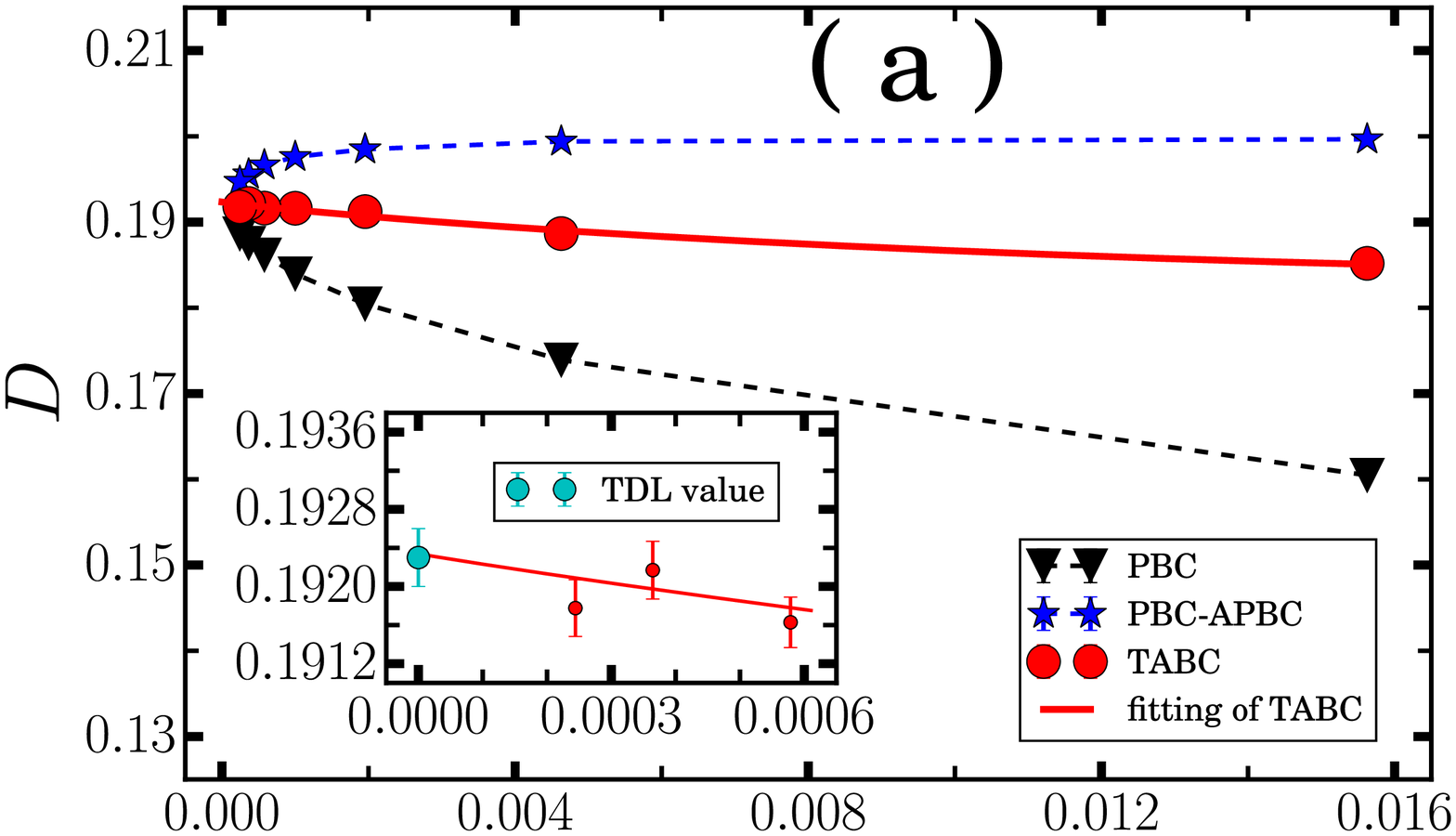}
	\includegraphics[width=8.0cm]{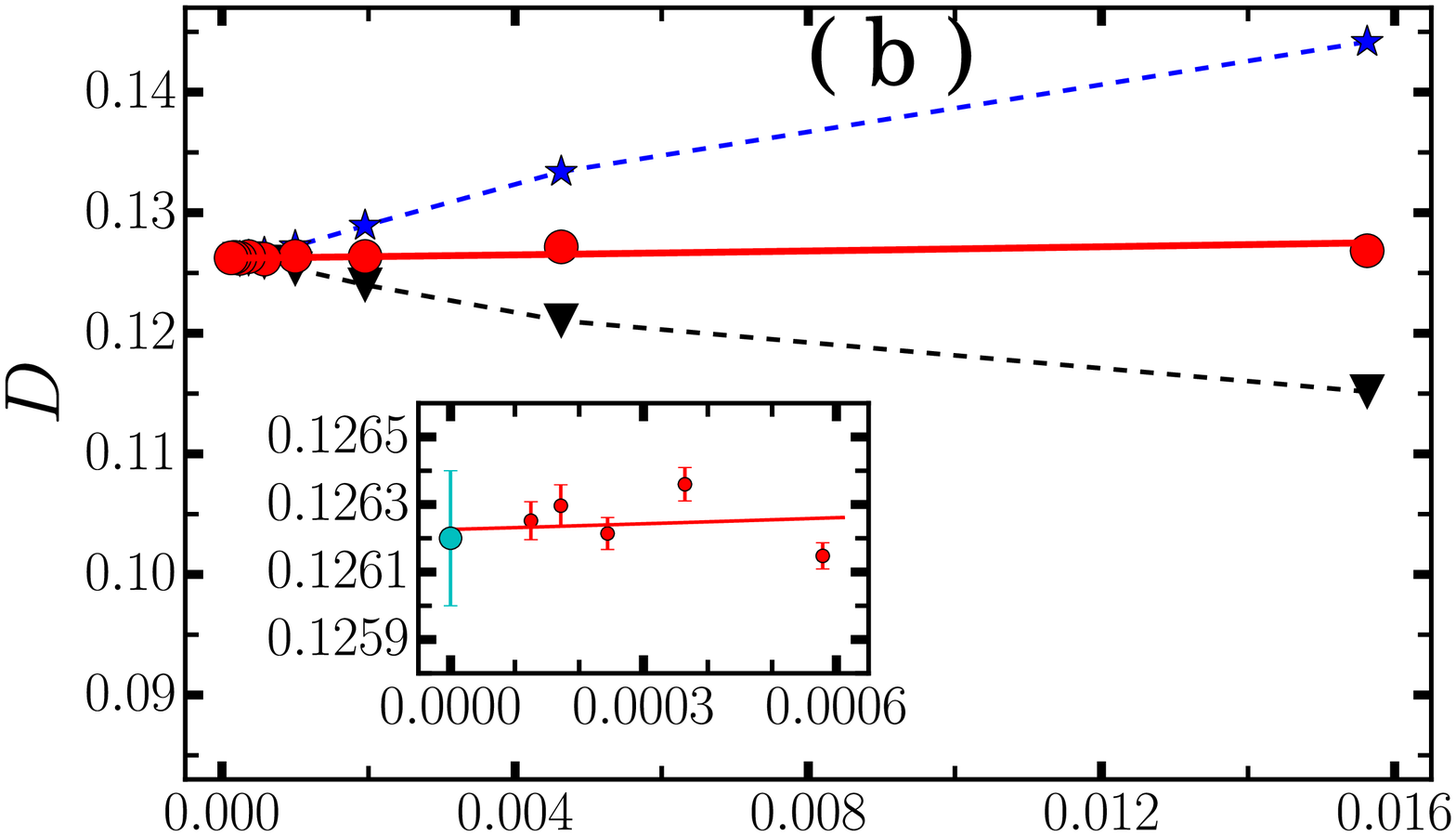}
	\includegraphics[width=8.0cm]{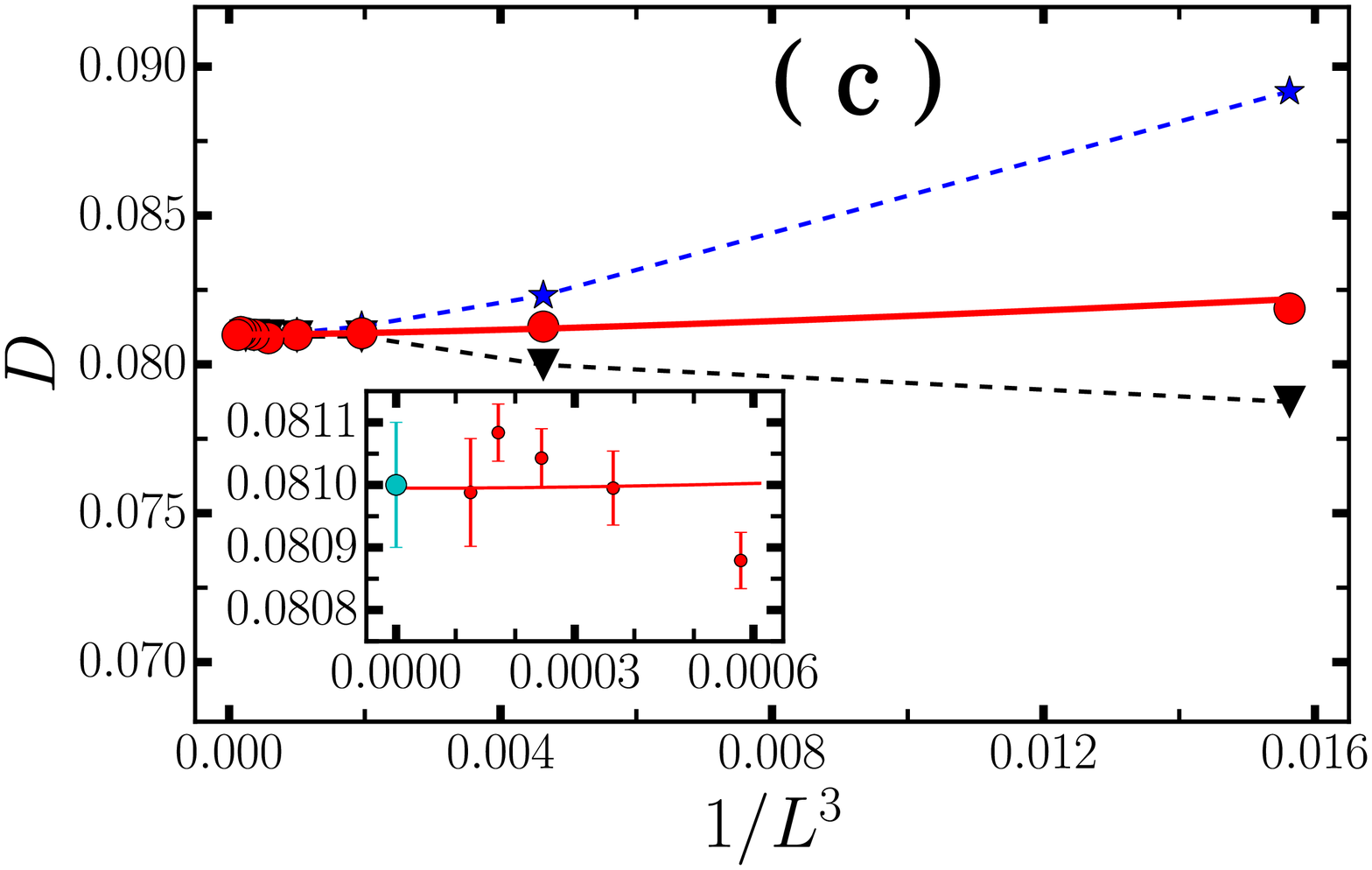}
	\includegraphics[width=8.0cm]{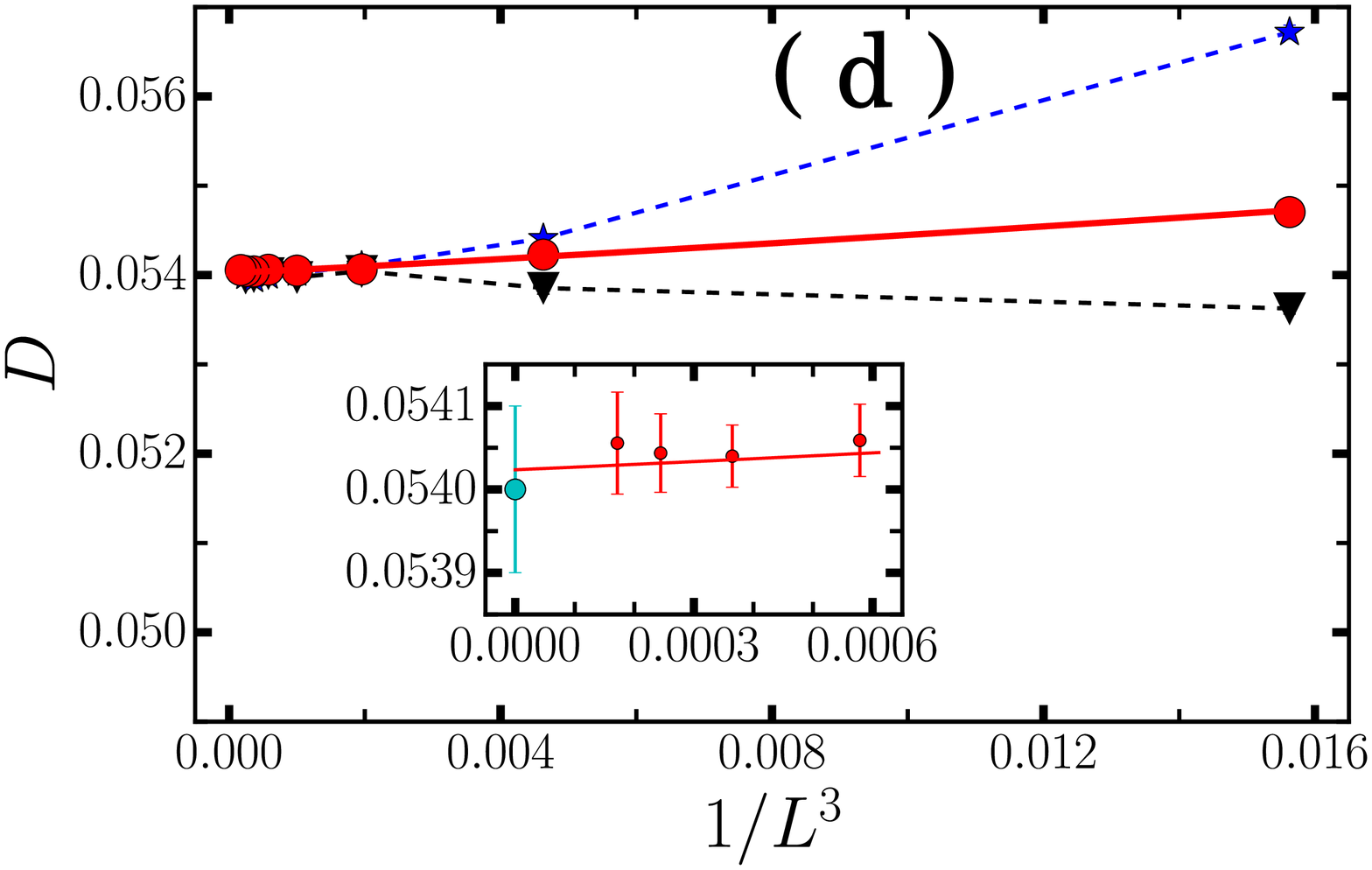}
	\caption{(Color online) Double occupancy at half-filling calculated using different boundary conditions. 
	Symbols and setup are similar to Fig.~\ref{E_half_all}.
		}
	\label{D_half_all}
\end{figure*}

\section{Results at half filling}
\label{results_half_filling}

In this section, we present results at half-filling. As mentioned, the AFQMC results are numerically exact,
as the sign problem is absent because of the particle-hole symmetry. 
We use a combination of the 
path-integral approach \cite{FG2D-Hao} and the random walk approach \cite{lecture-notes}.
With the former, an infinite variance problem exists which make the Monte Carlo error bars unreliable 
and thus could render results from standard AFQMC calculations incorrect \cite{Hao-inf-var}. The infinite variance
problem was removed \cite{Hao-inf-var}
 in our calculations, to obtain reliable results and error estimates on the observables.
Results are presented for the ground state energy, double occupancy, effective hopping, and staggered magnetization for $U = 2, 4, 6,$ and $8$. Detailed finite-size data are given, up to $16 \times 16$, to provide benchmarks for  future theoretical and computational studies. Careful extrapolation and analysis are then performed to obtain
results at the thermodynamic 
limit from the finite-size data. 
   
\subsubsection{Energy, Double Occupancy, and effective hopping}

We consider three types of boundary conditions here, i.e. PBC, PBC-APBC, and TABC. Here PBC-APBC means periodic along the $x$ direction and anti-periodic along the $y$ direction, which gives a closed-shell at half-filling.
In Fig.~\ref{E_half_all}, we plot the ground state energies versus supercell size
for all three boundary conditions. Detailed data are given in Appendix~\ref{apdix_E_D_finite}. 
As seen in the table there, our PBC and PBC-APBC data typically range from $4\times4$ to $16\times 16$.
Our TABC data contain about 200 twists for the smaller supercells to about 6 twists for $20\times 20$. 
The statistical error bars contain joint QMC and twist uncertainties. 
The fits to reach the TDL are also shown in Fig.~\ref{E_half_all}, with the insets 
displaying the asymptotic regime with the TABC, from which the TDL values are obtained.

Our fit for the ground-state energy has the following form:
\begin{equation}
E_0/L^2 = e_0 + a/L^3 +b/L^4
\label{scaling_E}
\end{equation}
where $e_0$ is the energy per site at the TDL. 
In the large $U$ limit at half-filling, the Hubbard model reduces to the spin-$1/2$ Heisenberg model with coupling constant $J = 4t^2 / U$
\cite{MacDonald_prb_1988}. From spin density wave theory, the leading order of finite size correction of ground state energy per site for the latter
is $1/L^3$ on a square lattice \cite{finite_size_scaling_E}. This scaling relationship was also confirmed by quantum Monte Carlo calculations \cite{sandvik_1997}. 
Our scaling choice in Eq.~(\ref{scaling_E}), based on these considerations, is seen to fit the data in the Hubbard model with excellent accuracy.

From Fig.~\ref{E_half_all} we see that the TABC energies tend to lie between the PBC and PBC-APBC results.
With PBC and PBC-APBC, the curves are less smooth. In fact the PBC energies are non-monotonic
for $U = 4$ and $U = 6$. 
To enter the scaling region of Eq.~(\ref{scaling_E}), large system size is needed, 
which makes extrapolation to the TDL challenging.
 The finite size effect is reduced with TABC, as expected from our discussion in the previous section. 
 Even at small system sizes, the scaling relationship in Eq.~(\ref{scaling_E}) holds well,
 making the fit more robust
 comparing to that using PBC and PBC-APBC data. 
 With a least squares fit
 of the TABC data, a reliable estimate of the ground state energy in TDL is obtained. 
 For $U = 2, 4, 6$, and  $8$, the final ground state energies per site
 are $-1.1760(2)$, $-0.8603(2)$, $-0.6567(3)$, and $-0.5243(2)$, respectively. 
 (The ground state energy for $U = 4$ is consistent with a previous QMC result $-0.85996(5)$
obtained with a $45$ degree tilted supercell \cite{sandro_U4_result})

 The magnitude of the finite size effect is seen to decrease with $U$. 
 (Note the vertical scales are different in the different panels.)
 This is the result of a balance of one-body and two-body finite-size effects.
 The one-body effects are especially pronounced at low $U$ because of shell effects. 
  The two-body finite-size effects are weakened in the Hubbard model because of the 
 very short-range nature of the interaction. 
 That the TABC results fit the ansatz in Eq.~(\ref{scaling_E}) so well across the entire range of 
 lattice sizes for all interactions is an indication of the separation (or additive nature) of the 
 one- and two-body finite-size effects. 
 The relative improvement of TABC over other boundary conditions is the largest at low $U$.
 At large $U$, the effect of boundary condition is suppressed, 
 and the finite-size effect is dominated by the interaction and the antiferromagnetic correlation.
 All three boundary conditions give results that fall on the same finite-size curve of 
  Eq.~(\ref{scaling_E}) for lattice sizes beyond $L\sim 8$

In Fig.~\ref{D_half_all}, we plot the double occupancy, $D = \langle \sum_i n_{i\uparrow} n_{i\downarrow}\rangle/N$.  
Similar to the situation with the ground state energy, the data with TABC
 lie between the PBC and PBC-APBC data and the finite size effect is reduced by using TABC. We carry out a least squares fit of 
the TABC data using the 
scaling relationship given in Eq.~(\ref{scaling_E}), although the variation with $L$ is not large
compared to the statistical error bars, and the extrapolation is insensitive to the precise form 
used here.
The TDL value obtained by
the fits are $0.1923(3)$, $0.1262(2)$, $0.0810(1)$, and $0.0540(1)$ 
for $U = 2$, $4$, $6$, and $8$, respectively. The double occupancy decreases rapidly with
$U$ as expected. 

\begin{figure}[t]
	\includegraphics[width=8.0cm]{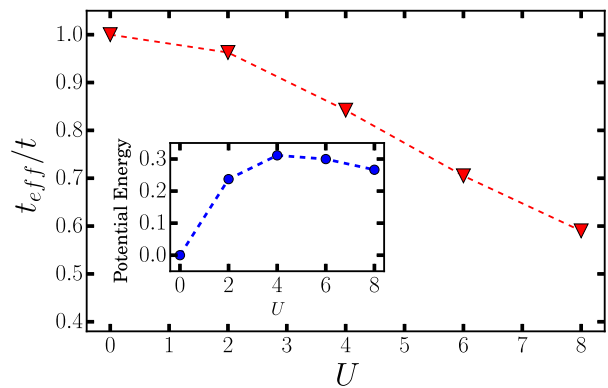}
	\caption{(Color online) The dependence of the effective hopping, $t_{\rm eff} / t$, on the interaction strength $U$ at half-filling. 
	The inset shows the corresponding potential energy in units of the non-interacting kinetic energy.
	}
	\label{t_eff_n1}
\end{figure}

To help quantify
the effect of $U$ on the bandwidth, we calculate the effective hopping $t_{\rm eff}/t$ \cite{white_U4_result} , defined as the 
ratio of kinetic energy in the presence of $U$ to
its non-interacting ($U = 0$) value,
\COMMENTED{
\begin{equation}
\frac{t_{\rm eff}}{t} = \frac{\langle c_{i\sigma}^\dagger c_j +  c_{j\sigma}^\dagger c_i \rangle _U}{\langle c_{i\sigma}^\dagger c_j +  c_{j\sigma}^\dagger c_i \rangle _0}
\end{equation}
}
\begin{equation}
\frac{t_{\rm eff}}{t} = \frac{\langle K \rangle _U}{\langle K \rangle_{U=0}}
\end{equation}
The kinetic energy can be obtained straightforwardly by subtracting 
the potential energy, given by $U$ times the double occupancy discussed above, from the total energy.
The effective hopping at the TDL is shown 
in Fig.~{\ref{t_eff_n1}} as a function of interaction.
The decrease of effective hopping with the increase of $U$ is consistent with
 the increasing of locality,
 as the system develops stronger antiferromagnetic order, which 
we characterize next.

We list the data of total energy, double occupancy, and kinetic energy with finite system size from $4 \times 4$ to $16 \times 16$ for PBC and PBC-APBC in Appendix~\ref{apdix_E_D_finite}.

\subsubsection{Spin correlations and magnetization}

\begin{figure}[b]
	\includegraphics[width=8.0cm]{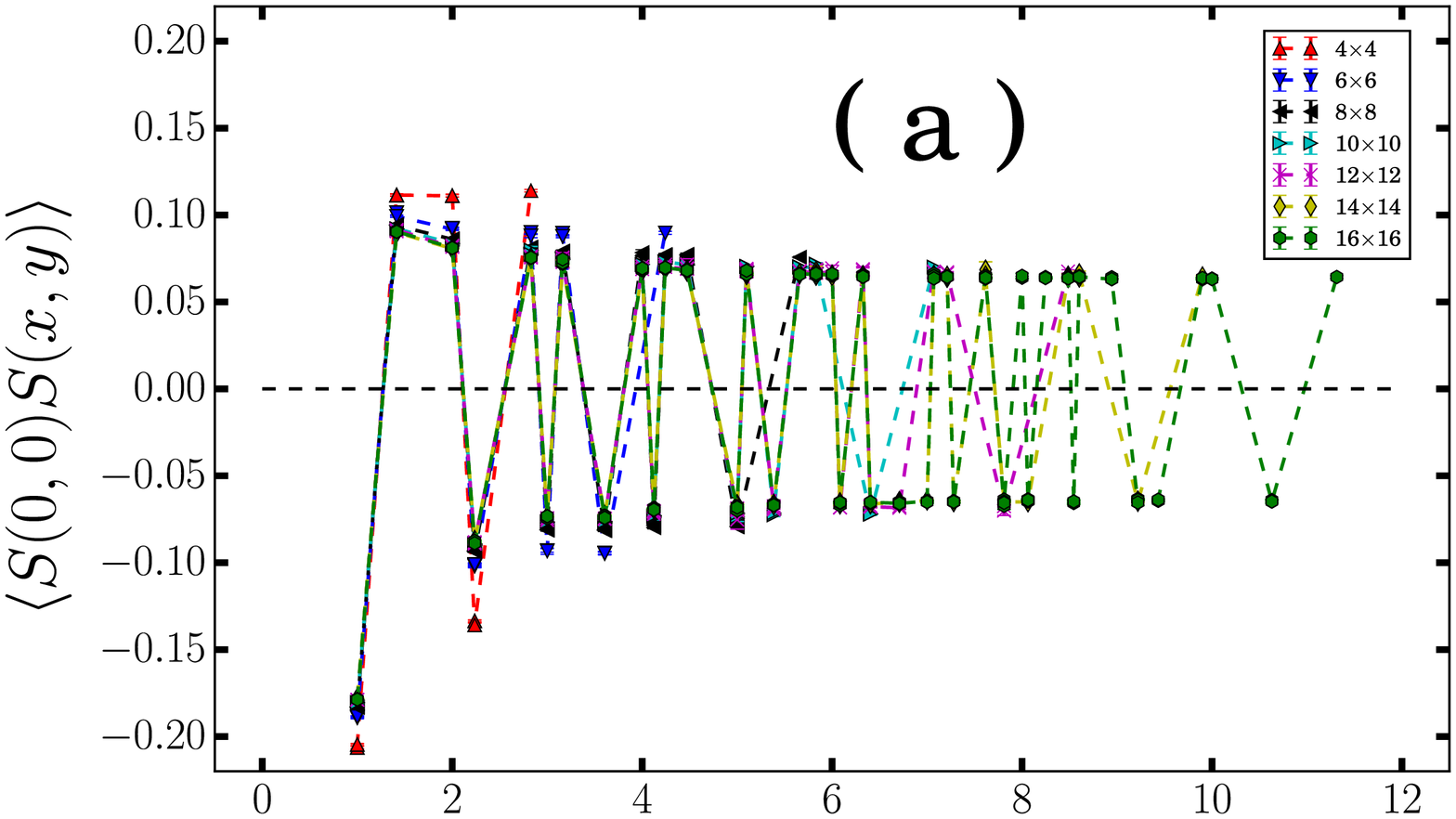}
	\includegraphics[width=8.0cm]{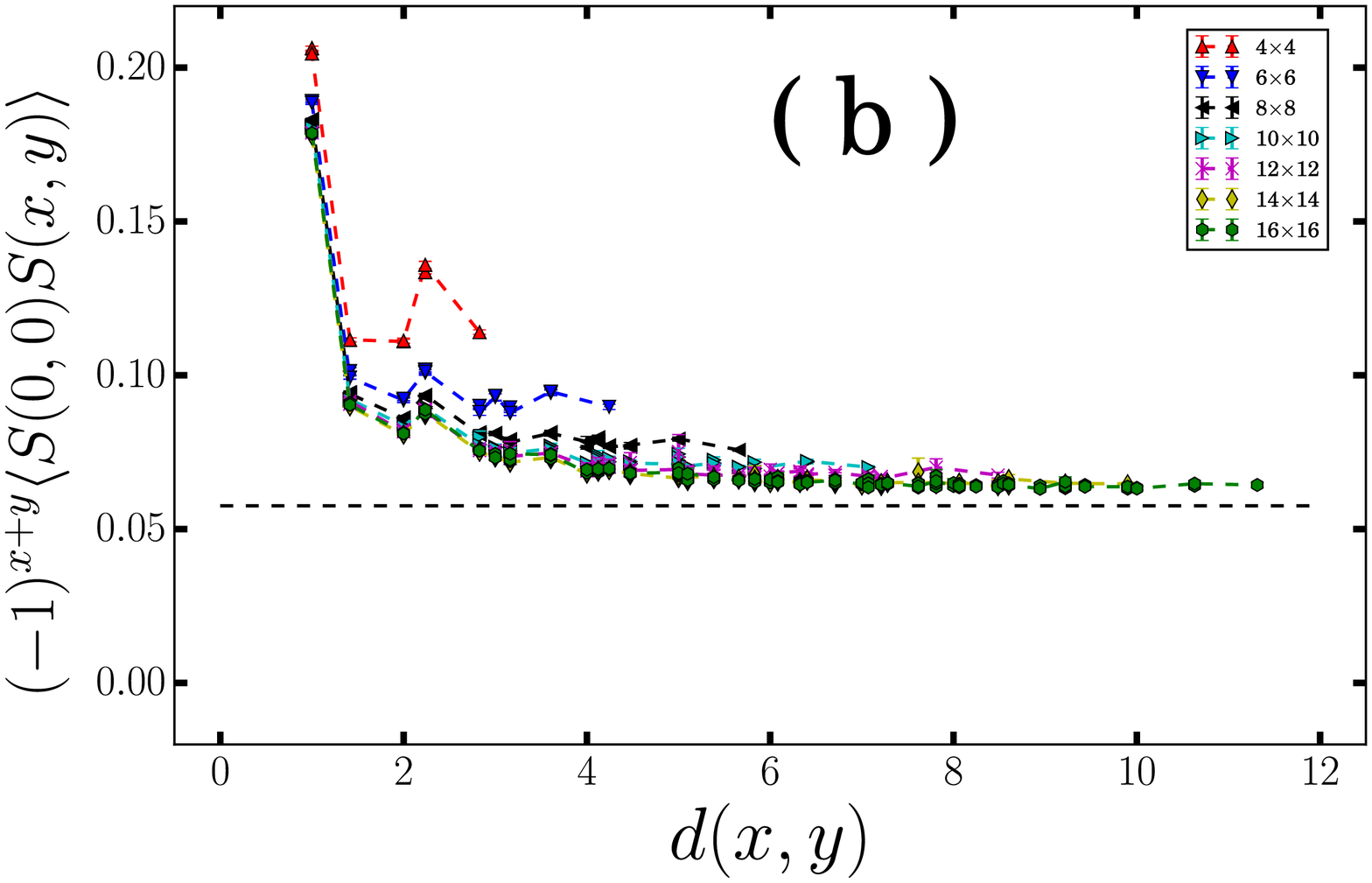}
	\caption{(Color online) Spin correlation function in the ground state at half filling. System sizes 
	ranging from $4 \times 4$ to $16 \times 16$ are shown, under PBC, with $U = 4$.
The horizontal axis is the relative distance, $\sqrt{x^2 + y^2}$. 
The top panel shows the spin correlation function, while the bottom panel shows the 
 staggered correlation. 
The dashed horizontal line in (b) shows the final TDL value obtained from the fit.
		}
	\label{spin-corr_U4_pbc}
\end{figure}

To quantify the magnetic properties in the ground state, we compute the spin correlation function,
 \begin{equation}
C(x, y) = \langle\psi_0|\textbf{S}(0,0)\cdot \textbf{S}(x,y)|\psi_0\rangle\,.
\label{def_corr}
\end{equation}
$\textbf{S}(x, y)$ is the spin operator at site $i$ with coordinate ($x$, $y$), which is given by
\begin{equation}
\textbf{S}(x,y)=\frac{1}{2}\sum_{ss^{\prime}}c_{is}^{\dagger}\overrightarrow{\sigma}c_{is^{\prime}}\,,
\end{equation}
where $\overrightarrow{\sigma}$ denotes the Pauli matrices.
In our calculation, translational symmetry is preserved statistically, so 
the reference point $(0,0)$ can be averaged over the whole lattice to reduce the statistical
error.
 In Fig.~\ref{spin-corr_U4_pbc}, we plot the ground-state spin correlation function for system sizes ranging from $4 \times 4$ to $16 \times 16$ under PBC 
for $U = 4$.
Long-range order is clearly seen. However, the strength of the correlation decreases
substantially from its short-distance values and also as system size is increased,  saturating to the asymptotic value very slowly with distance
and with system size.

\begin{figure}[t]
	\includegraphics[width=8.0cm]{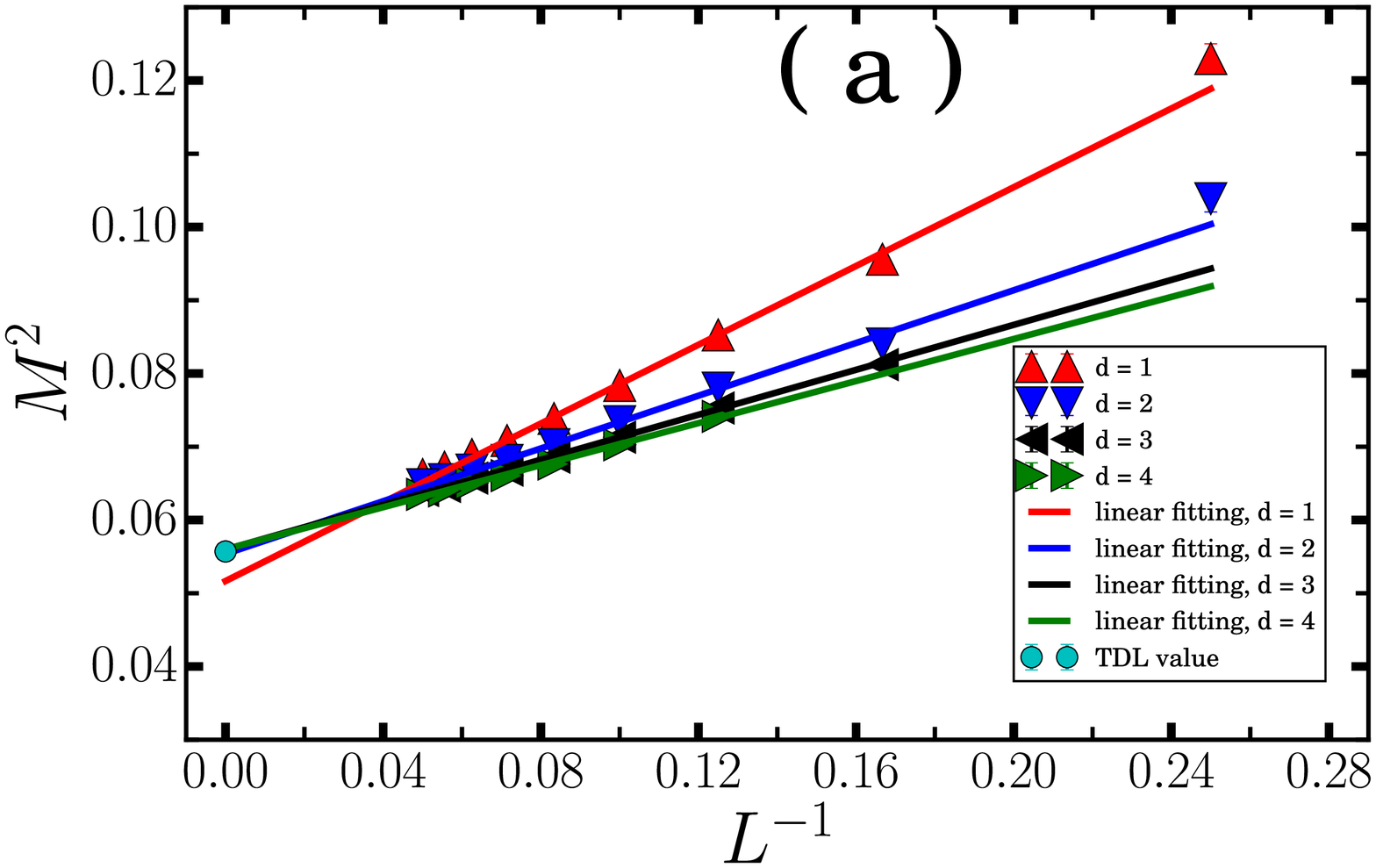}
	\includegraphics[width=8.0cm]{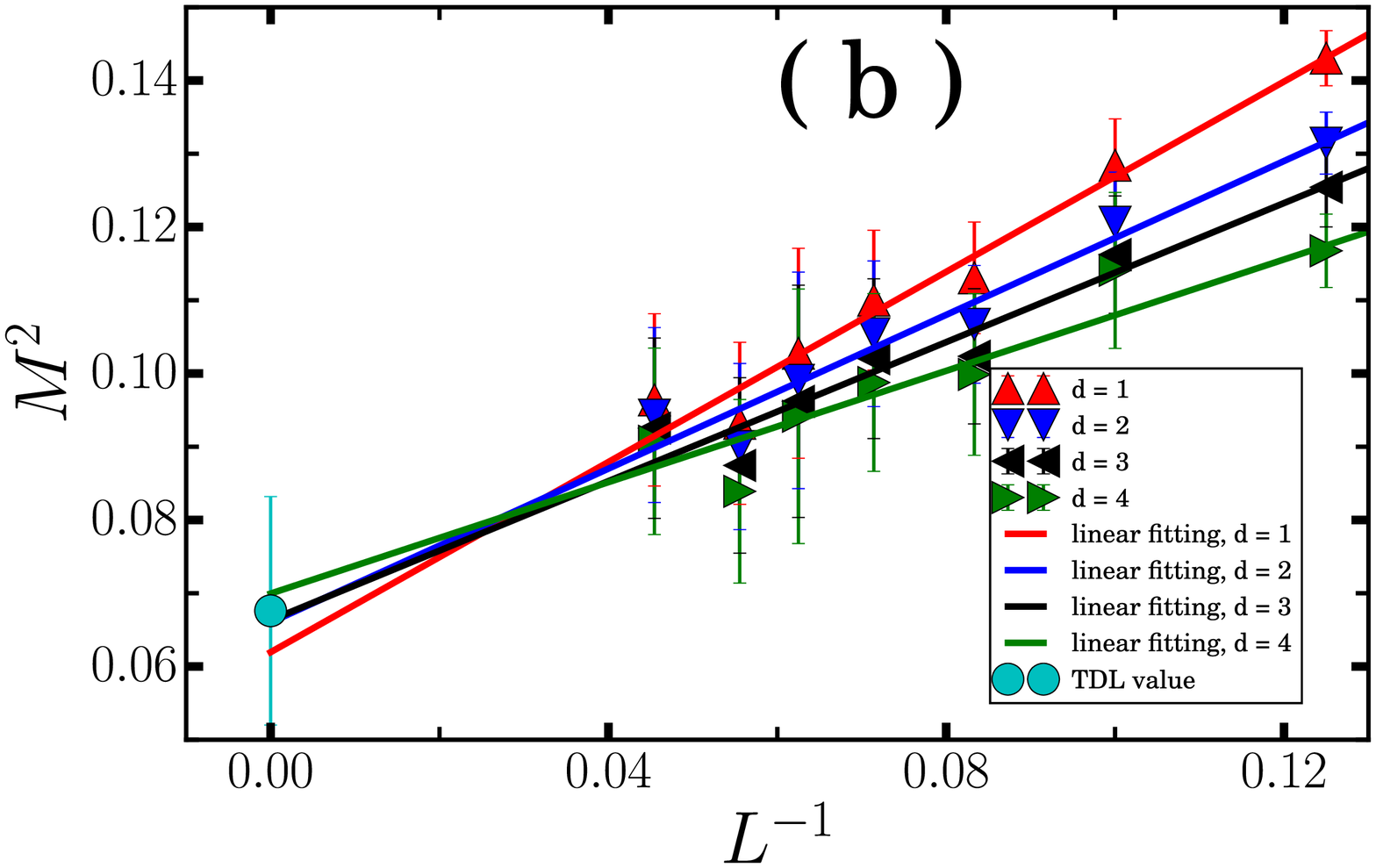}
	\caption{(Color online) Magnetization computed with TABC  at half-filling for (a) $U = 4$
	and (b) $U = 8$. For each choice of $d$, the result of a fit using the form in Eq.~(\ref{scaling_M}) 
	is also plotted. The cyan dot represents the final TDL value and the estimated error bar.
		}
	\label{M_half_all}
\end{figure}

We also compute the staggered magnetization. 
Two definitions
are usually used in the literature \cite{sandvik_1997}. One uses the spin-spin correlation function 
at the greatest distance which,
for a square lattice, is 
$M_1^2 = C(L/2, L/2)$.
The other
relies on the spin structure factor,
\begin{equation}
M_2^{2}=S(\pi,\pi)=\frac{1}{N}\sum_{i=1}^{N}(-1)^{x_{i}+y_{i}}C(x_{i},y_{i})\,.
\label{def_M_2}
\end{equation}
Both definitions have significant finite-size effects, as can be deduced from the results in  Fig.~\ref{spin-corr_U4_pbc}. 
 We use a modified definition \cite{scalettar_prb_2009}
\begin{equation}
 M(d)^{2}=\frac{1}{N-n}\sum_{x_{i}^{2}+y_{i}^{2}>d^2}^{N}(-1)^{x_{i}+y_{i}}C(x_{i},y_{i})\,,
 \label{def_M_d}
\end{equation}   
where  $n$ is the number of sites that fall within a sphere (circle) of radius $d$ centered at the 
reference point. 
All three definitions of the magnetization will converge to the same TDL value as $L\rightarrow \infty$. 
However, Eq.~(\ref{def_M_d}) gives a compromise which removes the large local effects near the reference point while averaging over multiple distances of the long-range correlation to reduce fluctuations. 

The computed magnetizations are plotted 
in Fig.~\ref{M_half_all} for $U = 4$ and $8$. 
In each case, we show results for a sequence of choices for $d$. 
We fit the computed magnetization as a function of supercell size, for each choice of $d$, with the following
scaling form 
\begin{equation}
M^{2}=M_{0}^{2}+\frac{a}{L}+O(\frac{1}{L^{2}})\,,
\label{scaling_M}
\end{equation}
where $M_0$ is the staggered magnetization at the TDL. 
Similar to scaling forms used above, the form in Eq.~(\ref{scaling_M}) is motivated by spin-wave theory \cite{finite_size_scaling_M}.
The evolution of the fitting with  $d$ is illustrated 
in the figure. 
The TDL results of magnetizations
are $0.119(4), 0.236(1), 0.280(5)$, and $0.26(3)$ for $U = 2, 4, 6$, and $8$, 
respectively. Note that the U = 2 result is different from that
listed in Ref.~\cite{paper_simons} which contained an error in the extrapolation
to the TDL.
An upper bound for the magnetization is given by the value of $0.3070(3)$,
from 
the spin-$1/2$ Heisenberg model on a square lattice \cite{sandvik_1997}. 
Our results are consistent with the scenario that the long-range AFM order persists to small $U$ values, 
with no Mott transition at finite $U$ in the two-dimensional Hubbard model at half-filling.

\section{Results away from half-filling}
\label{result_away_half-filling}

We next study the ground state when the system  is doped. The constrained-path approximation is applied to control the 
sign problem, as mentioned.
Previous studies have shown that the systematic error from the constraint in the CPMC calculation is 
small in the Hubbard model  \cite{chia-chen_EOS}.
We carried out additional benchmarks to further quantify the systematic errors \cite{paper_simons}.
At low and intermediate densities, the CP errors are small, using free electron TWFs. At higher densities where
magnetic correlation is enhanced, the GHF trial wave function improves the CP result and brings them to 
a level roughly comparable to that at intermediate densities, as discussed in Sec.~\ref{ssec:GHF}.

All results reported in this work have thus used single-determinant TWFs. 
Recent progress has resulted in further improvement in the accuracy of CPMC,
by use of symmetry properties \cite{CPMC_sym_1,CPMC_sym_2}, by constraint release \cite{CPMC_sym_1},
or by 
improving the trial wave function 
within CPMC via a self-consistent iteration \cite{Mingpu-sc-cpmc}. 
We have used  multideterminant trial wave functions and constraint release
to verify the accuracy in a few systems of larger $L$. The results are consistent with the benchmark discussed
above.

\subsubsection{Low to medium density}

\begin{figure}[h]
	\includegraphics[width=4.2cm]{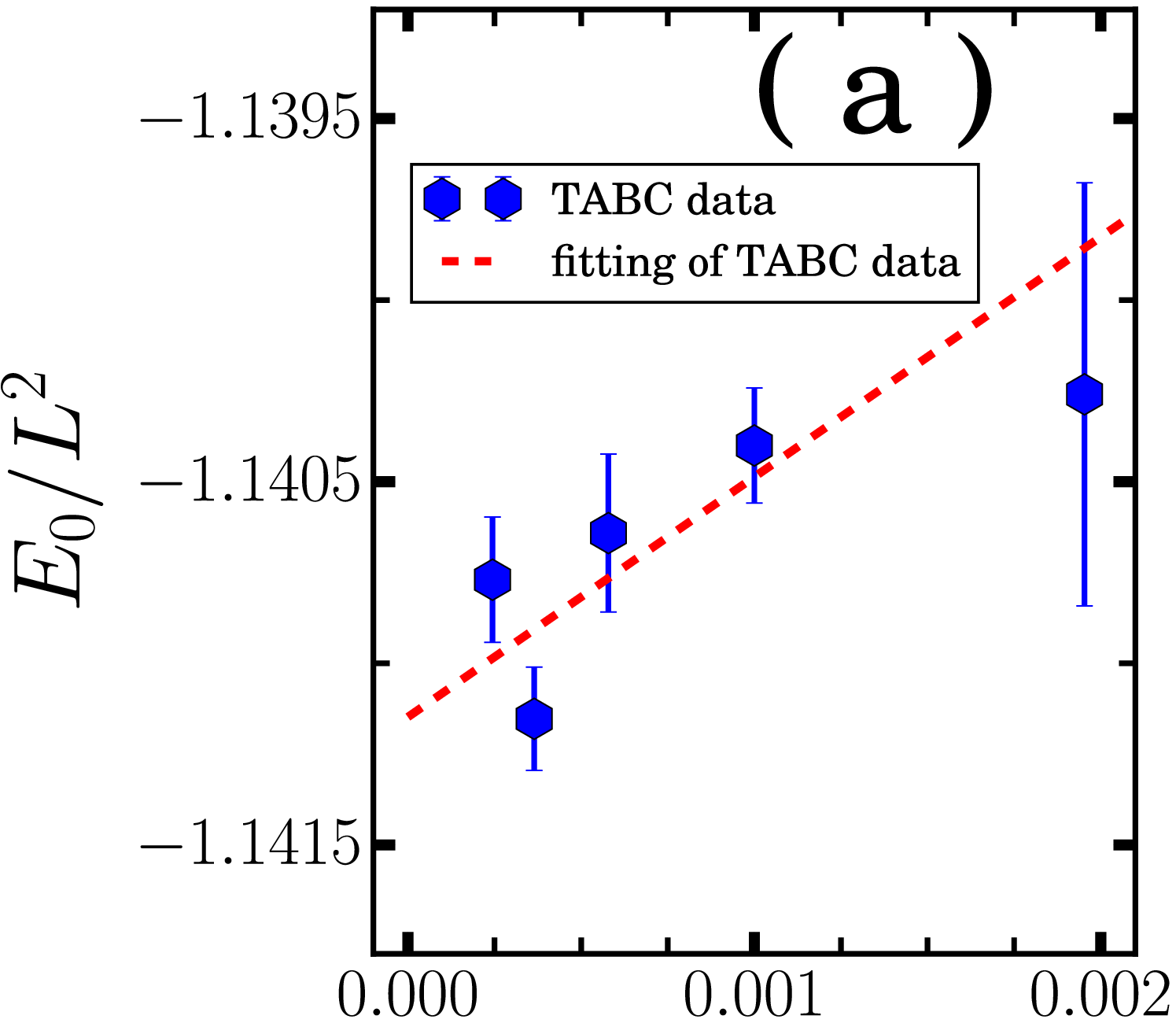}
	\includegraphics[width=4.0cm]{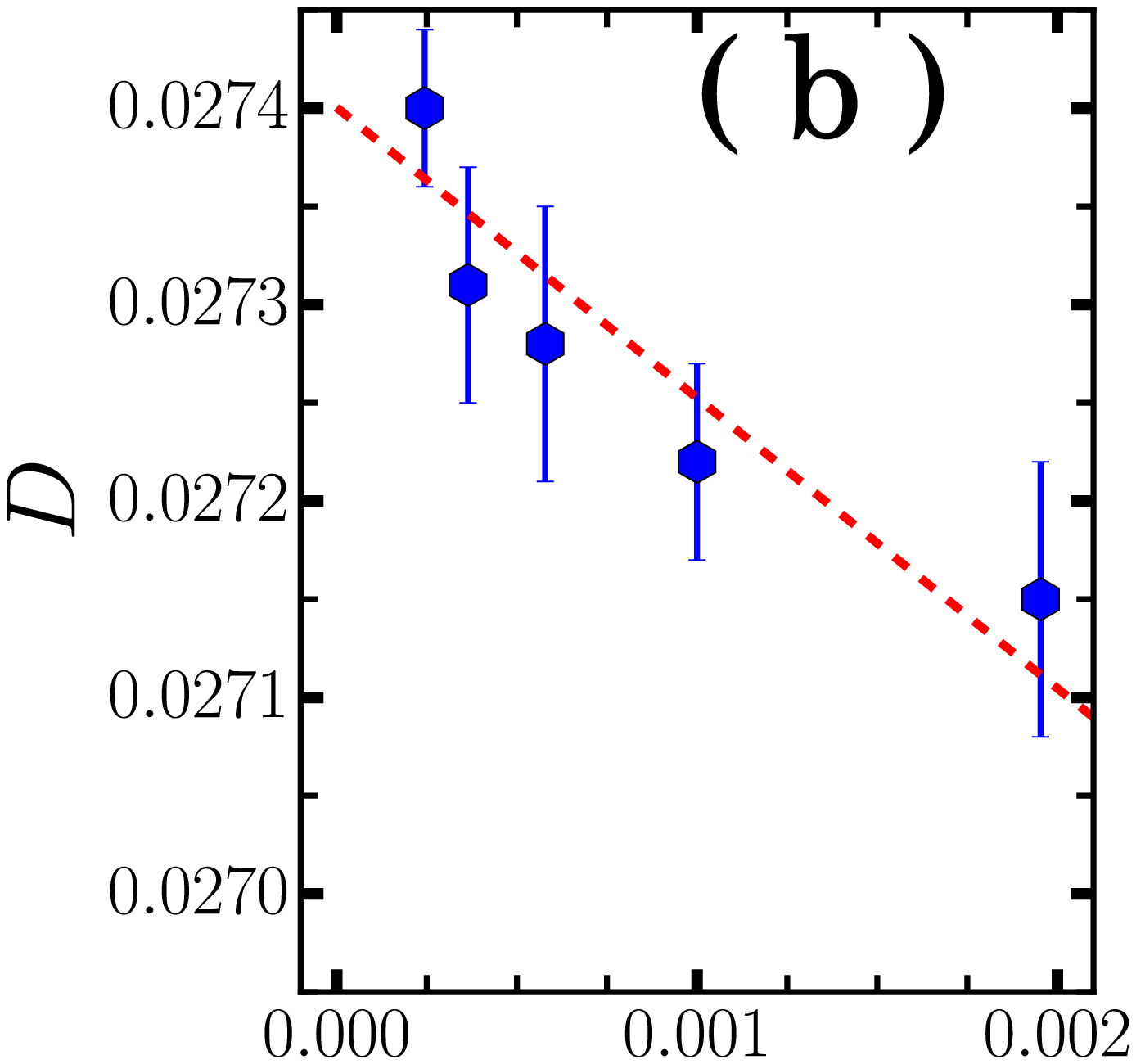}
	\includegraphics[width=4.2cm]{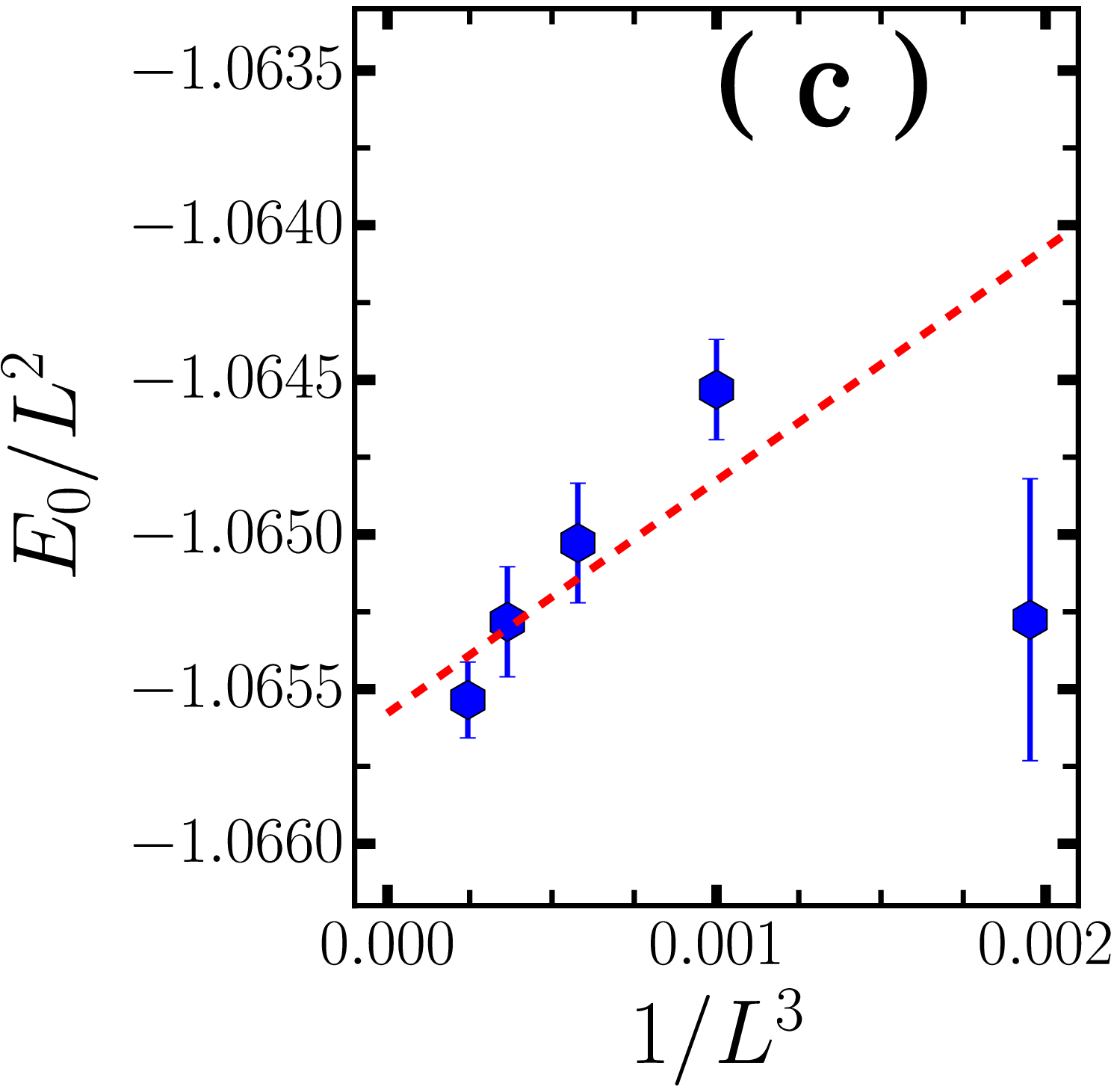}
	\includegraphics[width=4.0cm]{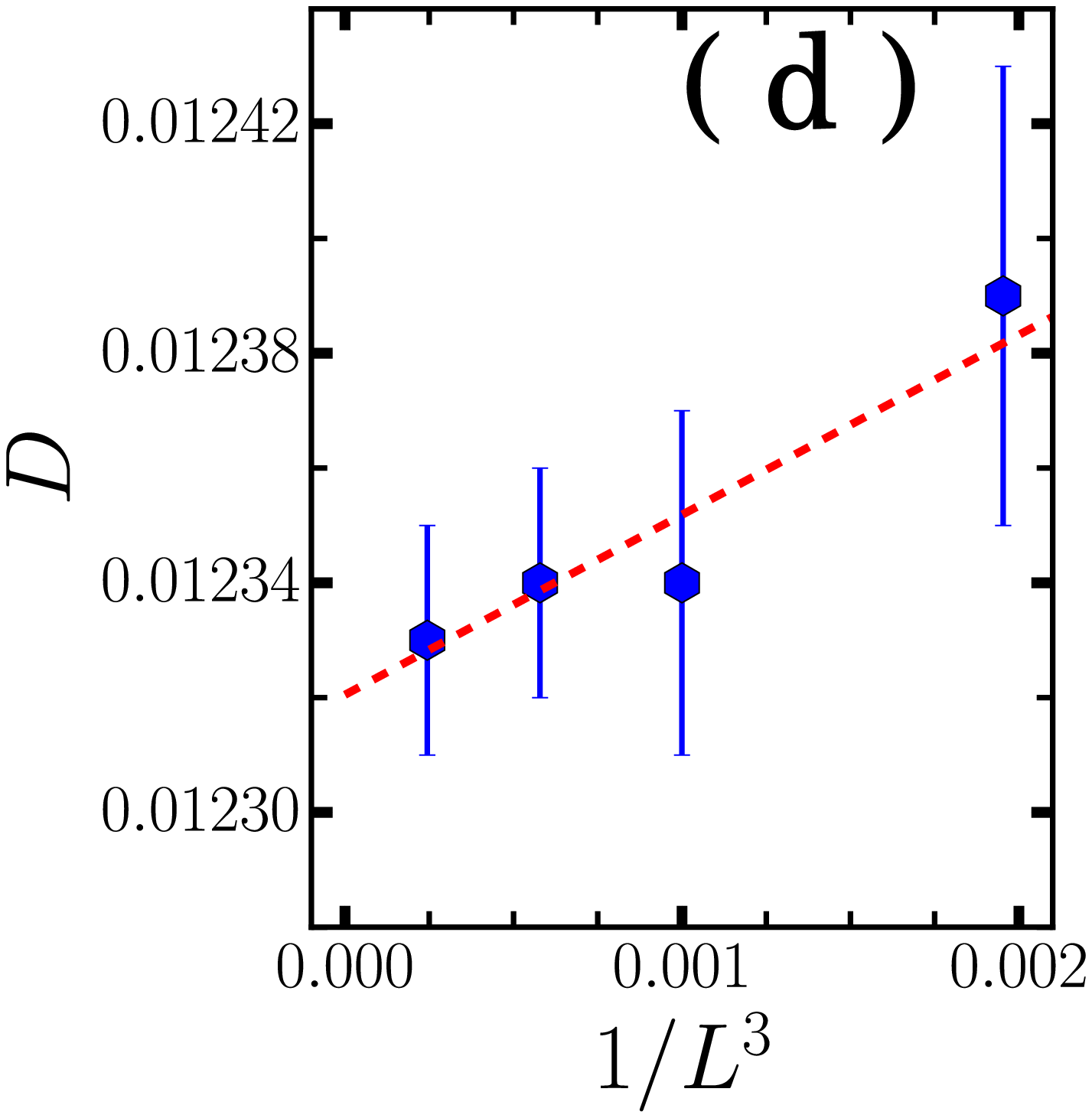}
	\caption{(Color online) Ground-state energy and double occupancy vs.~supercell size at $n = 0.5$ for $U = 4$ and $U = 8$. TABC is used.
		Panels (a) and (b) correspond to $U = 4$. Panels (c) and (d) correspond to $U = 8$. 
	The solid lines are from a fit using  $E_0 / L^2 (D) \sim e_0 (D_0) + a / L^3$.  
	}
	\label{E_n0.5}
\end{figure}

In this section, we present numerical results for densities of $n=0.3$, $0.5$, $0.6$, and $0.75$ in the 
TDL. 
We first illustrate the finite-size effects and the extrapolation to the TDL with $n=0.5$, which can be precisely 
realized for any even $L$.  In Fig.~\ref{E_n0.5} (a) and (c), we plot the ground-state energy for $U = 4$ and $U = 8$,
using TABC. The corresponding double occupancy is presented in Fig.~\ref{E_n0.5} (b) and (d).
We have also relaxed the targeted statistical accuracy somewhat compared to half-filling,
because of CP systematic errors. Given this and given the large system 
sizes we compute,
the residual finite-size effects are modest.
For example, the results from $16\times 16$ lattices with TABC are indistinguishable from the 
extrapolated TDL value within statistical errors.   
Both quantities are seen to continue to  fit well the general form in Eq.~(\ref{scaling_E}), being linear 
 in $1/L^3$ for large $L$. 
With double occupancy, the TABC reduces the finite-size effects substantially. The residual two-body 
finite-size effects are seen to have opposite slopes for $U=4$ and $U=8$. Similar behavior is seen 
in the results at half-filling presented in Fig.~\ref{D_half_all}.

\begin{figure}[h]
	\includegraphics[width=8.0cm]{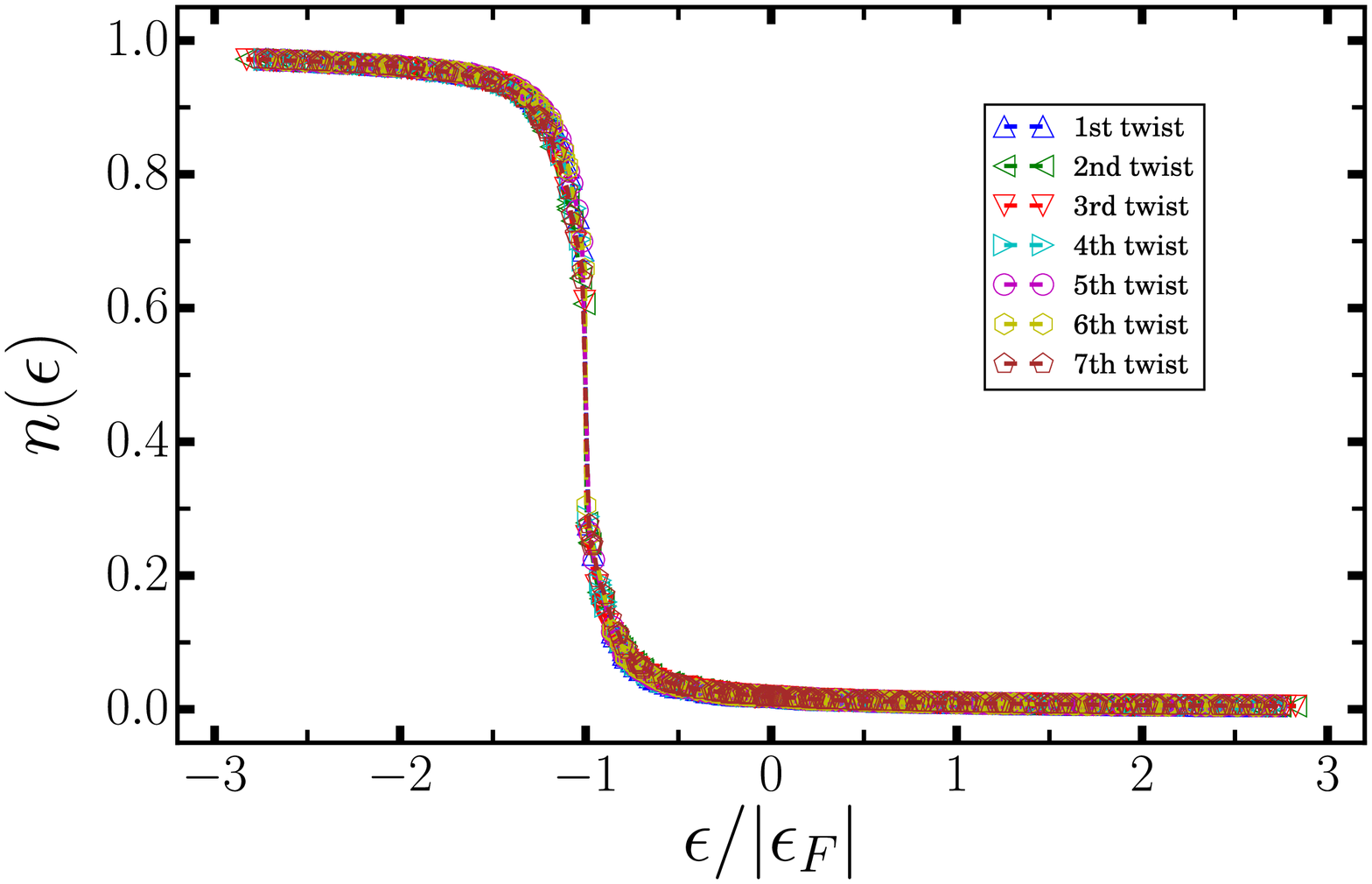}
	\includegraphics[width=8.0cm]{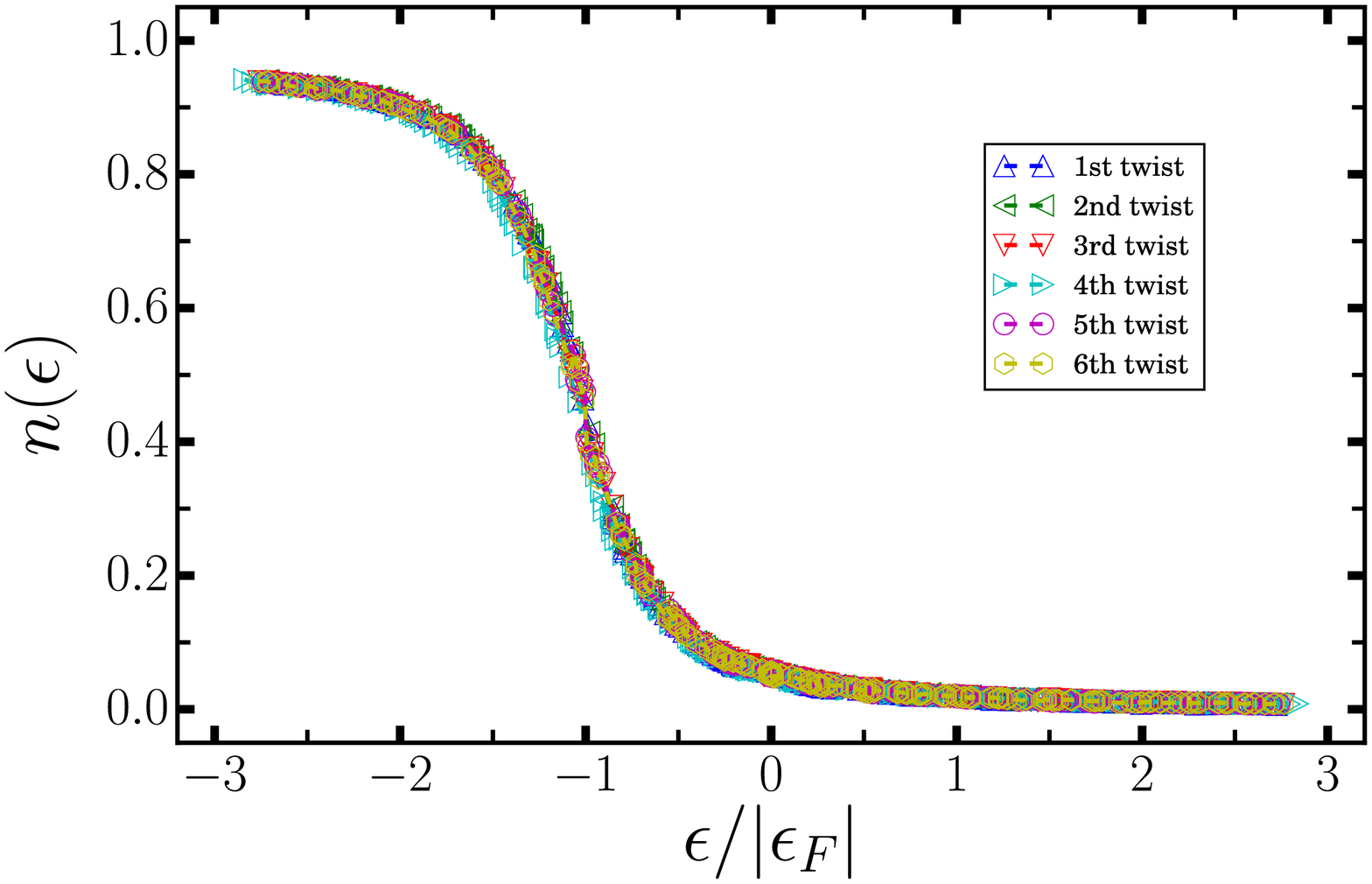}
	\caption{(Color online) Momentum distribution at $n = 0.5$ for (a) $U = 4$ and (b) $U = 8$. The horizontal axis is the 
		non-interacting energy for the given momentum normalized by the non-interacting Fermi energy of the corresponding twist. }
	\label{Mot-dis_n0.5}
\end{figure}

\begin{table}[h]	
	\centering
	\caption{ Ground state energy and kinetic energy per site, and double occupancy 
		for low to intermediate densities at $U=4$ and $U=8$. 
			}
	\begin{tabular}{c|ccccc}
		\hline
		\hline
		& $n$ & $0.3$ & $0.5$ & $0.6$ & $0.75$ \\ \hline
		\multirow{3}{*}{$U = 4$} & $e_0$& $-0.8793(2)$ & $-1.141(2)$& $-1.1845(5)$& $-1.1491(2)$  \\
		\hhline{~*{5}{-}}
 	     & $D$ & $0.00932(1)$& $0.02740(4)$& $0.0404(1)$& $0.06606(6)$ \\ 
 	     \hhline{~*{5}{-}}
	 	 & $k$ & $-0.9166(2)$ & $-1.251(2)$& $-1.3461(6)$& $-1.4133(3)$ \\ \hline
	 	
	 	\multirow{3}{*}{$U = 8$} & $e_0$& $-0.8534(1)$ & $-1.066(2)$& $-1.0729(1)$& $-0.9666(4)$  \\
	 	\hhline{~*{5}{-}}
	 	& $D$ & $0.00442(1)$& $0.01232(2)$& $0.01776(3)$& $0.02847(4)$ \\ 
	 	\hhline{~*{5}{-}}
	 	& $k$ & $-0.8888(1)$ & $-1.165(2)$& $-1.2150(3)$& $-1.1944(5)$ \\
	 	\hline
	 	\hline
	\end{tabular}
	\label{E_D_away-half}
\end{table}

Similar calculations and analysis were carried out for the other densities. 
For $n = 0.3$ and $0.6$, integer fillings are not possible in certain finite systems. In these cases, we
interpolate from the results for the nearest two integer fillings. 
 A prior study \cite{chia-chen_EOS} had computed the equation of state for $U=4$. Our results 
in this density range are consistent with theirs. 
In Table~\ref{E_D_away-half} we list
the ground-state energies, double occupancies, and kinetic energies for all densities studied in this regime for both $U=4$ and $U=8$. 

We also computed the momentum distribution at $n = 0.5$ which is shown in Fig.~\ref{Mot-dis_n0.5}.
For each $U$ we plot the results for several twist angles. The $x$ axis is the 
non-interacting energy for the given momentum normalized by the non-interacting Fermi energy of the corresponding twist.   
For $U = 4$, we can find a a obvious discontinuity, which is a indicator of the Fermi liquid behavior in this system and agree with an early QMC
calculation\cite{moreo_prb_1990}. For $U = 8$, there is no obvious jump.

\subsubsection{$n = 0.875$}

The nature of the ground state at  $n = 0.875$ is still not completely known. 
Many competing tendencies are present including spin density wave, charge density wave, and possibly superconducting order \cite{fradkin_rmp}.
We did not measure the superconducting correlation function in this work.
(Prior calculations with
CPMC using free-electron trial wave functions did not find long-range pairing correlation 
in the ground state with the resolution possible then \cite{pairing1997}.) 
 In a previous study \cite{chia-chen_prl}, a spin density wave (SDW) ground state with 
wave length $\lambda = 16$ ($2/h$) was found at $n = 0.875$ and $U = 4$.
The computed energies with supercells which are commensurate with the SDW wavelength are
seen to be slightly lower than those which are not.
Our new GHF trial wave functions gave results consistent with this.

To accommodate the SDW structure, we 
studied a range of systems with sizes $4 \times 16, 8 \times 16,
16 \times 16, 8\times 32$, and $8\times 48$. The energy per site under TABC for these were $-0.7674(7)$, $-0.7658(3)$, $-0.7657(2)$, $-0.7657(4)$, and $-0.7660(3)$, respectively, at $U = 8$.
The energies are consistent with each other except for the one with the smallest width of $4$.
 A conservative estimate the ground state energy in the TDL is $-0.766(1)$. 
Similarly, the TDL value for $U = 4$ 
is estimated to be $-1.026(1)$. [The corresponding energy results using free-electron trial wave-functions
are $-0.773(1)$ and $-1.032(1)$, respectively. Based on the analysis and benchmark discussed earlier, these
are expected to be less accurate than the GHF results.]
The corresponding double occupancy values are $0.0403(2)$ 
for $U=8$ and $0.0940(3)$ for $U=4$.

\section{Conclusion}
\label{conclusion}

The Hubbard model is one of the most fundamental models in many-body physics. It is often used
as a test ground as new approaches are developed in the quest to reliably treat interacting fermion 
systems or correlated materials.
In this work we have presented detailed benchmark results for the ground state of the two-dimensional Hubbard model. 
The total energy, double occupancy, effective hopping, spin correlation function, and magnetization are computed with the AFQMC method. 

At half-filling,
the results
are numerically exact. By a finite size scaling of the TABC data, 
the most accurate values to date  of these quantities are obtained. We also provide the finite size data
for system sizes ranging from $4 \times 4$ to $16 \times 16$ so as to facilitate benchmark of future 
analytical and computational studies. 
 
 Away from half-filling, we employ the constrained path CPMC method, which removes the sign problem 
 and allows us to systematically reach large system size in the same manner as at half-filling.
 Prior results and a new set of benchmark calculations here  show that the systematic error from
 the constraint is small. 
Results are presented from low to intermediate densities for $U/t=4$ and $8$. 
We also study the case of $n=0.875$ with a new form of single Slater determinant trial wave function,
obtaining energetics and determining the spin correlations for both values of $U/t$.

In addition to the generalized Hartree-Fock trial wave functions, 
which we have shown to  improve the accuracy of the constraint, 
we have also introduced
the use of quasi random twist sequences when implementing twist boundary conditions.
The quasi random twists
allow convergence with the number of twists which is  as fast
as
a uniform grid, while
eliminating any shell effects from degeneracies in the single-particle levels. 
The connection between GHF and BCS trial wave functions, and their interplay with 
the form of Hubbard-Stratonovich transformations will have broader impacts beyond Hubbard models.

\begin{acknowledgments}
We acknowledge the Simons Foundation for funding. SZ
and HS acknowledge support from NSF (DMR-1409510)
for AFQMC method development. MQ was also supported
by DOE (\MakeUppercase{DE-SC}0008627). We thank G. Chan and B.X.
Zheng for valuable exchanges and C.-C. Chang for a careful
reading of the manuscript. Earlier calculations were carried
out at the Oak Ridge Leadership Computing Facility at the
Oak Ridge National Laboratory. Most of the calculations
were carried out at the Extreme Science and Engineering
Discovery Environment (XSEDE), which is supported by
National Science Foundation Grant No. ACI-1053575, and the
computational facilities at the College of William and Mary.

\end{acknowledgments}


\begin{thebibliography}{1}

\bibitem{Hubbard_origional}
J. Hubbard, Proceedings of the Royal Society of London
A: Mathematical, Physical and Engineering Sciences {\bf 276}, 238 (1963).

\bibitem{Imada_rmp_1998}
Masatoshi~Imada, Atsushi~Fujimori, and Yoshinori~Tokura,
Rev. Mod. Phys. {\bf 70}, 1039(1998).

\bibitem{cdw_sdw}
Jie~Xu, Chia-Chen~Chang, Eric~J.~Walter, Shiwei~Zhang,
J. Phys.: Condens. Matter {\bf 23}, 505601 (2011), 
Robert~Peters and Norio~Kawakami, Phys. Rev. B {\bf 89}, 155134(2014) 

\bibitem{ph-sym}
J.~E.~Hirsch, Phys. Rev. B {\bf 31}, 4403 (1985).

\bibitem{htc}
P. W. Anderson, Science {\bf 235}, 1196 (1987),
Elbio~Dagotto, Rev. Mod. Phys. {\bf 66}, 763 (1994).

\bibitem{lieb_wu}
E. H. Lieb and F. Y. Wu, Phys. Rev. Lett. {\bf 20}, 1445 (1968).

\bibitem{paper_simons}
J. P. F. LeBlanc, Andrey E. Antipov, Federico Becca, Ireneusz W. Bulik, Garnet Kin-Lic Chan, Chia-Min Chung, Youjin Deng, Michel Ferrero, Thomas M. Henderson, Carlos A. Jimenez-Hoyos, E. Kozik, Xuan-Wen Liu, Andrew J. Millis, N. V. Prokofev, Mingpu Qin, Gustavo E. Scuseria, Hao Shi, B. V. Svistunov, Luca F. Tocchio, I. S. Tupitsyn, Steven R. White, Shiwei Zhang, Bo-Xiao Zheng, Zhenyue Zhu, and Emanuel Gull, Phys. Rev. X {\bf 5}, 041041(2015).


\bibitem{VMC-DMC-GFMC}
L. F. Tocchio, F. Becca, A. Parola, and S. Sorella, Phys. Rev. B {\bf 78}, 041101(R) (2008);
L. F. Tocchio, F. Becca, and C. Gros, Phys. Rev. B {\bf 83}, 195138 (2011).
N. Trivedi and D. M. Ceperley, Phys. Rev. B {\bf 41}, 4552 (1990).

\bibitem{MPHF}
 R. Rodriguez-Guzman, C. A. Jimenez-Hoyos, R. Schutski,
 and G. E. Scuseria, Phys. Rev. B {\bf 87}, 235129 (2013)
 
\bibitem{CC}
R. J. Bartlett and M. Musial, Rev. Mod. Phys. {\bf 79}, 291 (2007).
 
\bibitem{DMRG}
S. R. White, Phys. Rev. Lett. {\bf 69}, 2863 (1992).
 
\bibitem{DMET}
Gerald Knizia and Garnet Kin-Lic Chan, Phys. Rev. Lett. {\bf 109}, 186404 (2012).

\bibitem{DCA}
T. Maier, M. Jarrell, T. Pruschke, and M. H. Hettler, Rev. Mod. Phys. {\bf 77}, 1027 (2005).

\bibitem{DF}
A. N. Rubtsov, M. I. Katsnelson, and A. I. Lichtenstein, Phys. Rev. B {\bf 77}, 033101 (2008).

\bibitem{diagMC}
N. Prokofev and B. Svistunov, Phys. Rev. Lett. {\bf 99}, 250201 (2007).


\bibitem{sign}
E.~Y.~Loh Jr., J.~E.~Gubernatis, R.~T.~Scalettar, S.~R.~White, D.~J.~Scalapino, and R.~L.~Sugar, Phys. Rev. B {\bf 41}, 9301 (1990).
 
\bibitem{sign_problem}
Matthias Troyer and Uwe-Jens Wiese, Phys. Rev. Lett. {\bf 94}, 170201 (2005).
 
\bibitem{TBC}
C.~Lin, F.~H.~Zong and D.~M.~Ceperley, Phys. Rev. E {\bf 64}, 016702 (2001).

\bibitem{chia-chen_EOS}
Chia-Chen Chang and Shiwei Zhang, Phys. Rev. B {\bf 78}, 165101 (2008).

\bibitem{cheong_thesis}
Siew~Ann~Cheong, Ph.D. Dissertation, Cornell University (2006).

\bibitem{Chiesa-FS}
Simone Chiesa, David M. Ceperley, Richard M. Martin, and Markus Holzmann, Phys. Rev. Lett.{\bf 97}, 076404 (2006).

\bibitem{Hendra-FS}
Hendra Kwee, Shiwei Zhang, and Henry Krakauer, Phys. Rev. Lett. {\bf 100}, 126404 (2008).

\bibitem{sandro_prb_2015}
Sandro~Sorella, Phys. Rev. B {\bf 91}, 241116 (2015) .

\bibitem{AFQMC}
R.~Blankenbecler, D.~J.~Scalapino and R.~L.~Sugar, Phys. Rev. D {\bf 24}, 2278 (1981); G. Sugiyama and S. E. Koonin, Ann. Phys. (N.Y.) {\bf 168}, 1 1986.

\bibitem{lecture-notes}
S. Zhang, Auxiliary-Field Quantum Monte Carlo for Correlated Electron Systems, Vol. 3 of Emergent Phenomena in Correlated Matter: Modeling and Simulation, Ed. E. Pavarini, E. Koch, and U. Schollw¨ock (Verlag des Forschungszentrum J¨ulich, 2013).

\bibitem{hirsch_prb_1983}
J.~E.~Hirsch,
Phys. Rev. B {\bf 28}, 4059 (1983).

\bibitem{CPMC_sym_1}
Hao Shi and Shiwei Zhang, Phys. Rev. B {\bf 88}, 125132 (2013).

\bibitem{CPMC_sym_2}
Hao Shi, Carlos A.~Jimenez-Hoyos, R.~Rodriguez-Guzman, Gustavo E.~Scuseria, and Shiwei Zhang, Phys. Rev. B {\bf 89}, 125129 (2014).

\bibitem{Hao-inf-var}
Hao Shi and Shiwei Zhang, Phys. Rev. E {\bf 93}, 033303 (2016).

\bibitem{zhang_prb_1997}
S.~Zhang, J.~Carlson, and J.~E.~Gubernatis, Phys. Rev.
B {\bf 55}, 7464 (1997).

\bibitem{Wirawan-PRE}
Wirawan Purwanto and Shiwei Zhang, Phys. Rev. E {\bf 70}, 056702 (2004).

\bibitem{Huy_CPC_2014}
Huy Nguyen, Hao Shi, Jie Xu, Shiwei Zhang, Comput. Phys. Commun {\bf 185}, 3344 (2014).

\bibitem{halton}
J.~H.~Halton, Commun. ACM {\bf 7}, 701 (1964).

\bibitem{GHF}
Sharon~Hammes~Schiffer and Hans~C.~Andersen, J. Chem. Phys. {\bf 99}, 1901 (1993).

\bibitem{Phaseless}
S.~Zhang and H.~Krakauer, Phys. Rev. Lett. {\bf 90}, 136401 (2003).

\bibitem{small_twist}
For open shell system, we can use TBC with a small twist to break the degeneracy.

\bibitem{Wirawan-F2-spin-contamination}
Wirawan Purwanto, W. A. Al-Saidi, Henry Krakauer, Shiwei Zhang, J. Chem. Phys. {\bf 128}, 114309 (2008).

\bibitem{BCS_wave-f}
J. Carlson, Stefano Gandolfi, Kevin E. Schmidt, and Shiwei Zhang, Phys. Rev. A {\bf 84}, 061602 (2011).

\bibitem{Scalettar_1989}
R. T. Scalettar, E. Y. Loh, J. E. Gubernatis, A. Moreo, S. R. White, D. J. Scalapino, R. L. Sugar, and E. Dagotto, Phys. Rev. Lett. {\bf 62}, 1407 (1989).

\bibitem{FG2D-Hao}
Hao Shi, Simone Chiesa, Shiwei Zhang, Phys. Rev. A {\bf 92}, 033603 (2015).

\bibitem{MacDonald_prb_1988}
A.~H.~MacDonald, S.~M.~Girvin and D.~Yoshioka, Phys. Rev. B {\bf 37}, 9753 (1988).

\bibitem{finite_size_scaling_E}
Herbert Neuberger and Timothy Ziman, Phys. Rev. B {\bf 39}, 2608 (1989), Daniel S.~Fisher, Phys. Rev. B {\bf 39}, 11783 (1989).

\bibitem{sandvik_1997}
Anders~W.~Sandvik, Phys. Rev. B {\bf 56}, 11678 (1997). 

\bibitem{sandro_U4_result}
Sandro~Sorella, Phys. Rev. B {\bf 84}, 241110 (2011).

\bibitem{white_U4_result}
S.~R.~White, D.~J.~Scalapino, R.~L.~Sugar, E.~Y.~Loh, J.~E.~Gubernatis, and R.~T.~Scalettar, Phys. Rev. B {\bf 40}, 506 (1989).

\bibitem{finite_size_scaling_M}
D.~A.~Huse, Phys. Rev. B {\bf 37}, 2380 (1988).

\bibitem{scalettar_prb_2009}
C.~N.~Varney, C.~R~Lee, Z.~J.~Bai, S.~Chiesa, M.~Jarrell, and R.~T.~Scalettar, Phys. Rev. B {\bf 80}, 075116 (2009)

\bibitem{Mingpu-sc-cpmc}
Mingpu Qin, Hao Shi, and Shiwei Zhang, in preparation. 

\bibitem{moreo_prb_1990}
A.~Moreo, D.~J.~Scalapino, R.~L.~Sugar, S.~R.~White, and N.~E.~Bickers, Phys. Rev. B {\bf 41}, 2313 (1990).

\bibitem{fradkin_rmp}
Eduardo~Fradkin, Steven~A.~Kivelson, and John~M.~Tranquada, Rev. Mod. Phys. {\bf 87}, 457 (2015).  

\bibitem{pairing1997}
Shiwei Zhang, J. Carlson, and J. E. Gubernatis, Phys. Rev. Lett. {\bf 78}, 4486 (1997).

\bibitem{chia-chen_prl}
C.-C.~Chang and S.~Zhang, Phys. Rev. Lett. {\bf 104}, 116402 (2010).
	
\end{thebibliography}

\appendix*
\section{Finite-size data at half-filling}
\label{apdix_E_D_finite}

In Table~\ref{QMC_ED_half_finite_ED}, we list the total ground-state energy, potential
energy, and kinetic energy with PBC and PBC-APBC.

\begin{table*}
	\centering
	\caption{Total ground state energy (E), potential energy (P), and kinetic energy (K) in the 
		Hubbard model at half-filling,  for $U = 2, 4, 6, 8$. Supercell cell sizes ranging from 
		$4 \times 4$ to $16 \times 16$ are studied. Results are listed for both PBC and PBC-APBC.
		Statistical errors are on the last digit and are indicated in parenthesis.
	}
		\begin{tabular}{c|ccc|cc|cc|cc}
			\hline
			\hline  
		&	& \multicolumn{2}{c|}{ $U = 2$} & \multicolumn{2}{c|}{ $U = 4$} & \multicolumn{2}{c|}{ $U = 6$} & \multicolumn{2}{c}{ $U = 8$}\\ \hline
		 & & PBC & PBC-APBC & PBC & PBC-APBC & PBC & PBC-APBC  & PBC & PBC-APBC \\
		 	\hhline{~*{9}{-}}
		\multirow{3}{*}{$ 4 \times 4$} & $E$ & -18.024(6) & -20.114(2) & -13.616(6) & -14.594(3) & -10.541(4) & -10.902(7) & -8.476(9) & -8.646(8)  \\
		\hhline{~*{9}{-}}
		& $P$ & 5.135(9) & 6.389(1) & 7.370(8) & 9.227(7) & 7.559(4) & 8.56(1) & 6.864(8) & 7.26(1) \\ 
		\hhline{~*{9}{-}}
		& $K$ & -23.164(3) & -26.503(3) & -20.989(4) & -23.823(8) & -18.097(7) & -19.46(1) & -15.34(2) & -15.90(2) \\ \hline
		
		\multirow{3}{*}{$ 6 \times 6$} & $E$& -41.457(5) & -43.499(2) & -30.865(9) & -31.43(2) & -23.74(1) & -23.84(1) & -19.00(2) & -19.01(1)  \\
		\hhline{~*{9}{-}}
		& $P$ & 12.522(7) & 14.357(1) & 17.43(1) & 19.21(2) & 17.276(7) & 17.78(1) & 15.51(2) & 15.67(1) \\ 
		\hhline{~*{9}{-}}
		& $K$ & -53.982(4) & -57.857(4) & -48.230(8) & -50.63(1) & -41.012(9) & -41.62(1) & -34.51(2) & -34.69(2) \\ \hline
		
		\multirow{3}{*}{$ 8 \times 8$} & $E$& -74.470(5) & -76.308(3) & -55.05(1) & -55.31(1) & -42.16(2) & -42.17(2) & -33.68(3) & -33.66(2)  \\
		\hhline{~*{9}{-}}
		& $P$ & 23.098(5) & 25.407(6) & 31.747(8) & 33.006(9) & 31.08(2) & 31.21(1) & 27.67(2) & 27.70(2) \\ 
		\hhline{~*{9}{-}}
		& $K$ & -97.565(5) & -101.710(8) & -86.793(8) & -88.314(8) & -73.24(2) & -73.37(1) & -61.30(3) & -61.33(3) \\ \hline
		
		\multirow{3}{*}{$ 10 \times 10$} & $E$& -116.908(4) & -118.505(4) & -86.12(4) & -86.20(2) & -65.80(2) & -65.76(2) & -52.54(3) & -52.49(2)  \\
		\hhline{~*{9}{-}}
		& $P$ & 36.793(5) & 39.521(7) & 50.14(2) & 50.89(3) & 48.56(3) & 48.64(3) & 43.18(4) & 43.22(2) \\ 
		\hhline{~*{9}{-}}
		& $K$ & -153.699(5) & -158.024(9) & -136.23(1) & -137.10(2) & -114.37(2) & -114.41(2) & -95.73(3) & -95.69(3) \\ \hline
		
		\multirow{3}{*}{$ 12 \times 12$} & $E$ & -168.749(7) & -170.112(3) & -123.95(2) & -123.99(3) & -94.66(2) & -94.67(2) & -75.54(2) & -75.58(3)  \\
		\hhline{~*{9}{-}}
		& $P$ & 53.616(7) & 56.629(9) & 72.52(3) & 72.96(2) & 69.96(2) & 69.96(3) & 62.23(2) & 62.20(2) \\ 
		\hhline{~*{9}{-}}
		& $K$ & -222.364(8) & -226.741(9) & -196.48(1) & -196.95(2) & -164.64(2) & -164.64(2) & -137.77(4) & -137.74(4) \\ \hline
		
		\multirow{3}{*}{$ 14 \times 14$} & $E$& -229.981(6) & -231.134(4) & -168.67(2) & -168.69(3) & -128.76(2) & -128.78(3) & -102.85(3) & -102.83(4) \\
		\hhline{~*{9}{-}}
		& $P$ & 73.545(8) & 76.661(8) & 98.93(3) & 99.12(3) & 95.24(2) & 95.24(2) & 84.65(3) & 84.60(4) \\ 
		\hhline{~*{9}{-}}
		& $K$ & -303.530(6) & -307.795(8) & -267.59(2) & -267.82(2) & -224.01(2) & -224.02(3) & -187.49(4) & -187.47(4) \\ \hline
		
		\multirow{3}{*}{$ 16 \times 16$} & $E$& -300.596(6) & -301.562(5) & -220.29(4) & -220.30(4) & -168.19(3) & -168.21(5) & -134.23(3) & -134.25(3)  \\
		\hhline{~*{9}{-}}
		& $P$ & 96.585(8) & 99.694(7) & 129.16(4) & 129.36(5) & 124.36(3) & 124.33(6) & 110.53(3) & 110.57(3) \\ 
		\hhline{~*{9}{-}}
		& $K$ & -397.184(8) & -401.259(8) & -349.52(2) & -349.68(2) & -292.54(3) & -292.57(3) & -244.72(5) & -244.82(5) \\
		\hline
		\hline
	\end{tabular}
		\label{QMC_ED_half_finite_ED}
\end{table*}

\end{document}